\documentclass[11pt]{article}
\usepackage[margin=1in]{geometry}

\usepackage{amsmath,amssymb}
\usepackage{graphicx}
\usepackage{tikz}
\usetikzlibrary{arrows.meta,positioning,calc}
\usepackage{booktabs}
\usepackage{array}
\usepackage[colorlinks=true,linkcolor=blue,citecolor=blue,urlcolor=blue,pdfusetitle]{hyperref}
\newcommand{\dip}{\mathcal{D}}          
\newcommand{\den}{\mathsf{D}}           
\newcommand{\Nd}{\mathcal{N}}           
\newcommand{\En}{\mathcal{E}}           
\newcommand{\ERF}{E_{\mathrm{RF}}}
\newcommand{\ORF}{\Omega_{\mathrm{RF}}}
\newcommand{\dRF}{\Delta_{\mathrm{RF}}}
\newcommand{\GEIT}{\Gamma_{\mathrm{EIT}}}
\newcommand{\CEIT}{C_{\mathrm{EIT}}}
\newcommand{\Pdet}{P_{\mathrm{det}}}
\newcommand{\Rpd}{R_{\mathrm{pd}}}
\newcommand{\NEF}{\mathrm{NEF}}
\newcommand{\RIN}{\mathrm{RIN}}
\newcommand{\SFDR}{\mathrm{SFDR}}

\title{A Microwave-Photonics Formulation of the Rydberg-Atom Receiver}
\author{David Rosser\\
	The Aerospace Corporation\\
	\texttt{david.rosser@aero.org}}
\hypersetup{pdfauthor={David Rosser}}
\date{}

\begin{document}
\maketitle

\begin{abstract}
The performance of Rydberg-atom receivers is seldom stated as a radio-frequency (RF) link budget. The receiver is shown to be an intensity-modulated direct-detection analog optical link whose modulator is the laser-driven vapor. The transmission of a warm vapor is derived in closed form by averaging the four-level response over atomic velocities. The conversion gain, noise figure, bandwidth, intercept points, and spurious-free dynamic range follow from this one function. For a weak probe the instantaneous bandwidth is half the width of the transparency window in two-photon detuning. The local-oscillator beat places the second-order intercept near twice the local-oscillator field, so the local oscillator sets the even-order linearity. For a rubidium transition at 17.04~GHz the closed forms agree with a velocity-resolved density-matrix model to 1\% for the slope and the intercepts and to 10\% for the bandwidth. They underestimate the group delay by a third. The response is tied to atomic constants, and the sensing volume escapes the Chu limit. However, the noise figure of a warm vapor rules it out as a general replacement for an antenna and low-noise amplifier.
\end{abstract}
\section{Introduction}\label{sec:intro}

A Rydberg-atom receiver is an intensity-modulated direct-detection (IMDD)
analog optical link in which a laser-driven vapor takes the place of the
modulator. The identification is quantitative. The output voltage is the
photodiode current developed across a transimpedance, the current is the
detector responsivity multiplied by the received optical power, and the
received power is the launched probe power multiplied by the transmission
of the vapor, $T(\ERF)$. The conversion gain, the noise figure, the
instantaneous bandwidth, the intercept points, and the spurious-free
dynamic range (SFDR) of the receiver all follow from this one function,
which is obtained below in closed form for a warm vapor.

The transducer is the Rydberg electrometer
\cite{sedlacek_microwave_2012,fan_atom_2015,holloway_electric_2017}. A weak probe laser and a strong
coupling laser establish electromagnetically induced transparency (EIT) in a
warm alkali vapor, which opens a narrow transmission window on the probe. A
microwave field that couples a pair of Rydberg states splits that window
through the Autler--Townes (AT) effect \cite{autler_stark_1955}, and the
splitting returns the field in SI units because a known dipole matrix
element fixes it. With a radio-frequency (RF) local oscillator (LO) the
vapor acts as a
mixer and transfers a weak signal to an intermediate frequency (IF) on the
probe \cite{simons_rydberg_2019,gordon_weak_2019,jing_atomic_2020}.
Demonstrations span traceable field measurement from 1~GHz to 500~GHz
\cite{holloway_broadband_2014,gordon_millimeter_2014}, detuning-assisted
measurement of weak fields \cite{simons_using_2016}, and reception of
digitally modulated and broadcast waveforms
\cite{meyer_digital_2018,anderson_atomic_2021}.

\begin{figure}[t]
\centering
\begin{tikzpicture}[
    block/.style={rectangle, draw, thick, minimum width=2.8cm,
        minimum height=1.2cm, align=center, font=\sffamily},
    arrow/.style={-Stealth, thick, red!70!black},
    signal/.style={-Stealth, thick, green!60!black},
    lbl/.style={font=\small\sffamily, align=center},
    panel/.style={font=\large\bfseries\sffamily}
]
\node[panel, anchor=east] at (-1.8, 3) {a)};
\node[panel, anchor=east] at (-1.8, 0) {b)};
\node[block] (laser1) at (0, 3) {CW laser\\(carrier)};
\node[block, right=1.3cm of laser1] (mod) {modulator\\slope $\propto 1/V_\pi$};
\node[block, right=1.3cm of mod] (pd1) {photodiode\\$R_\text{pd}$};
\node[block, right=1.2cm of pd1, minimum width=1.8cm] (zt1) {$Z_T$};
\coordinate[right=0.8cm of zt1] (out1);
\draw[arrow] (laser1) -- (mod);
\draw[arrow] (mod) -- (pd1);
\draw[arrow] (pd1) -- (zt1);
\draw[arrow] (zt1) -- (out1) node[right, font=\normalsize] {$v$};
\draw[signal] ($(mod.south) + (0, -0.8)$) -- (mod.south);
\node[lbl, text=green!60!black, anchor=north] at ($(mod.south) + (0, -0.8)$)
    {RF in (wired)};
\node[block, below=3cm of laser1] (laser2) {probe laser\\(carrier)};
\node[block, right=1.3cm of laser2] (vapor) {vapor cell\\slope $\kappa_T$};
\node[block, right=1.3cm of vapor] (pd2) {photodiode\\$R_\text{pd}$};
\node[block, right=1.2cm of pd2, minimum width=1.8cm] (zt2) {$Z_T$};
\coordinate[right=0.8cm of zt2] (out2);
\draw[arrow] (laser2) -- (vapor);
\draw[arrow] (vapor) -- (pd2);
\draw[arrow] (pd2) -- (zt2);
\draw[arrow] (zt2) -- (out2) node[right, font=\normalsize] {$v$};
\draw[arrow, blue!70!black] ($(vapor.north) + (-0.9, 1.0)$)
    -- ($(vapor.north) + (-0.9, 0)$);
\node[lbl, text=blue!70!black, anchor=south] at ($(vapor.north) + (-0.9, 1.0)$)
    {coupling laser};
\draw[signal, dashed] ($(vapor.north) + (0.9, 1.0)$)
    -- ($(vapor.north) + (0.9, 0)$);
\node[lbl, text=green!60!black, anchor=south] at ($(vapor.north) + (0.9, 1.0)$)
    {RF LO};
\draw[signal] ($(vapor.south) + (0, -0.8)$) -- (vapor.south);
\node[lbl, text=green!60!black, anchor=north] at ($(vapor.south) + (0, -0.8)$)
    {$E_{\text{RF}}$ (free space)};
\end{tikzpicture}
\caption{The receiver as an analog optical link. (a) The intensity-modulated
direct-detection photonic link consists of a continuous-wave (CW) optical
carrier, an
electro-optic modulator biased at quadrature with slope $\propto1/V_\pi$,
and square-law detection into a transimpedance. (b) The Rydberg receiver
realizes the same three blocks. The probe laser is the carrier. The vapor is
the modulator, with transfer characteristic $T(\ERF)$, slope $\kappa_T$, and
a bias set at the EIT inflection by the RF local oscillator. The detection
chain is identical, and the output is the cascade $v=Z_T\Rpd P_0T$ of
Sec.~\ref{sec:cascade}. The RF enters from free space
(Sec.~\ref{sec:fieldmap}), and atomic constants fix the slope
(Sec.~\ref{sec:nf}).}
\label{fig:link}
\end{figure}
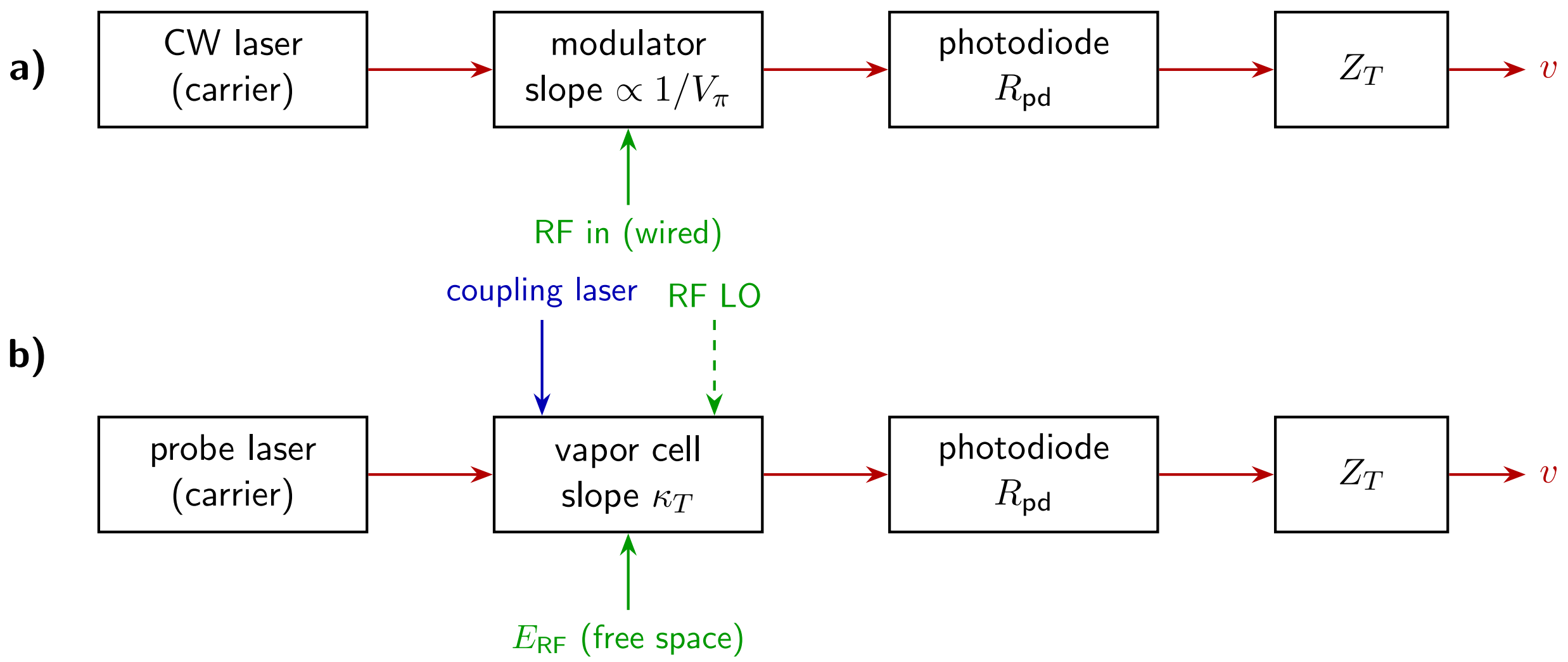

Figure~\ref{fig:link} draws the two signal chains block by block. In the
IMDD link of microwave photonics
\cite{urick_fundamentals_2015,cox_analog_2004,capmany_microwave_2007,marpaung_integrated_2019},
a continuous-wave laser provides an optical carrier, an electro-optic
modulator impresses the RF signal on the optical intensity, and a photodiode
recovers it by square-law detection. In the receiver the probe laser is the
carrier, the atomic medium is the modulator (its transmission is the
modulator transfer characteristic), and EIT with the LO establishes the
bias point and the modulation slope. The performance of an IMDD link is
governed by the optical power reaching the detector, the slope efficiency of
the modulator, and the noise of the detected carrier, and the same three
quantities govern the receiver. An unamplified photonic link has a noise
figure well above that of an electronic amplifier and an SFDR limited by
the curvature of the modulator
\cite{urick_fundamentals_2015,cox_analog_2004}. The receiver inherits both
limitations.

Two properties of the receiver have no counterpart in the photonic link and
motivate the device. Atomic constants fix the modulation slope, so the
receiver calibrates itself against the SI
\cite{sedlacek_microwave_2012,holloway_electric_2017}. The sensing element
is a sub-wavelength volume of vapor that neither radiates nor stores RF
energy as an antenna does, and the Chu size--bandwidth limit of an
electrically small antenna does not bind it
\cite{chu_physical_1948,wheeler_fundamental_1947}.

Each prior analysis of the receiver addresses part of the link budget. Fan
\emph{et al.} reviewed the vapor-cell sensor and the limits on its
sensitivity \cite{fan_atom_2015}. Cox \emph{et al.} introduced the standard quantum
limit for the data capacity of an electrically small receiver and observed
quantum-limited reception at bandwidths up to 30~MHz
\cite{cox_quantum-limited_2018}. Meyer \emph{et al.} calculated the
achievable sensitivity for field frequencies from 1~kHz to 1~THz with
analytical and semi-classical Floquet models, and compared the result with
electro-optic and antenna-coupled electronic sensors
\cite{meyer_assessment_2020}. Santamaria-Botello \emph{et al.} and Sandidge
\emph{et al.} translated measured sensitivities into an equivalent noise
temperature and concluded that free-space-coupled Rydberg receivers require
field enhancement to compete with room-temperature electronic front ends
\cite{santamaria-botello_comparison_2022,sandidge_resonant_2024}. Wu
\emph{et al.} modeled the harmonics and the linear dynamic range of the
superheterodyne receiver as a function of the probe Rabi frequency
\cite{wu_linear_2023}, and Gon\c{c}alves \emph{et al.} measured the
compression point, the intercept points, and the SFDR of one such receiver
\cite{goncalves_nonlinearities_2026}. However, these treatments do not
provide a single formulation from which the figures of merit of an RF link
budget, with the trade relations among them, follow from one function. A
two-pole response and closed-form expressions for the compression point and
the third-order intercept were also reported for a receiver based on
six-wave mixing \cite{chen_wideband_2026}. That analysis
does not address the second-order intercept, the noise figure, or the SFDR.

The microwave-photonics identification supplies that formulation. It imports
the link-budget formalism of the IMDD link, and every figure of merit
follows from the transmission function of the warm vapor, in closed form
for a weak probe. The Doppler-averaged transmission fixes the width and
depth of the transparency window and, through a lineshape constant, its
discriminator slope. The instantaneous bandwidth of a weakly probed vapor
is half the width of the window in the two-photon coordinate through which
the field moves every velocity class. The distortion is that of a cascade
of memoryless stages, in which the curvature of the absorption lineshape
sets the third-order intercept while the local-oscillator beat, with the
exponential of the Beer--Lambert law, sets the second-order intercept near
twice the local-oscillator field. The noise budget carries photon shot noise together with laser
intensity and frequency noise, and the laser terms set the practical floor
once the detected power is large, as in any direct-detection link. The closed forms are tested against a velocity-resolved numerical
model of the vapor (Appendix~\ref{app:numerics}), which also supplies the
effect of probe saturation, and against published measurements at their
reported operating points (Sec.~\ref{sec:validation}), with no fitted
parameter. Where the results overlap with prior analyses they agree. The
shot-noise-limited sensitivity is the photon-shot expression of
Refs.~\cite{cox_quantum-limited_2018,meyer_assessment_2020}, and the
aperture-referred noise figure reproduces the conclusion of
Refs.~\cite{santamaria-botello_comparison_2022,sandidge_resonant_2024} that a bare warm vapor
cannot compete with a low-noise amplifier without field gain ahead of the
cell.

Section~\ref{sec:model} derives $T$ and the small-signal model built on it.
Section~\ref{sec:sensdr} places the noise floor beneath that model and the
distortion ceiling above it, and translates both into system terms.
Section~\ref{sec:worked} evaluates one design and compares the model with
measurements. Appendix~\ref{app:atomic} collects the atomic constants that
parameterize $T$, and Appendix~\ref{app:numerics} describes the numerical
model.

\section{Small-signal model}
\label{sec:model}

The linear behavior of the receiver is set by a conversion gain from field
to voltage and by the bandwidth over which that gain holds. Both follow from
the transmission of the vapor.

\subsection{The receiver as a cascade}
\label{sec:cascade}

The back-end observable is the voltage across the transimpedance
$Z_T$ of the probe photoreceiver, and the photocurrent is the detector
responsivity times the received optical power,
\begin{equation}
v_{\mathrm{out}}(t) \;=\; Z_T\, i(t) \;=\; Z_T\,\Rpd\,\Pdet(t),
\qquad
\Rpd=\frac{\eta e}{h\nu_p},
\label{eq:current}
\end{equation}
with $\eta$ the detection quantum efficiency and $\nu_p$ the probe optical
frequency. The received power is the probe power $P_0$ launched into the vapor
cell multiplied by the transmission of the atomic medium,
\begin{equation}
\Pdet(t) \;=\; P_0\, T\!\big(\ERF(t);\,\nu\big),
\label{eq:Pdet}
\end{equation}
where the transmission depends on the applied RF field $\ERF$, on the optical
(two-photon) detuning $\nu$ at which the probe interrogates the medium, and,
through the finite response time of the atomic coherence, on the recent
history of that field. Chaining \eqref{eq:current} and \eqref{eq:Pdet},
\begin{equation}
v_{\mathrm{out}}(t) \;=\; Z_T\,\Rpd\,P_0\;T\!\big(\ERF(t)\big).
\label{eq:cascade}
\end{equation}
The atomic physics enters \eqref{eq:cascade} only through $T$, and the
analysis of the receiver is the analysis of this one function.

For a small field excursion $\delta E$ about an operating point, the output is
\begin{equation}
\begin{split}
\delta v
&= Z_T\,\Rpd\,P_0\,\frac{\partial T}{\partial \ERF}\,\delta E
= Z_T\,\Rpd\,\Pdet\;\kappa_T\,\delta E,\\
\kappa_T &\equiv \frac{1}{T}\frac{\partial T}{\partial \ERF},
\end{split}
\label{eq:kTdef}
\end{equation}
which defines the fractional transmission responsivity $\kappa_T$ (units
$(\mathrm{V/m})^{-1}$), the atomic analog of a modulator slope efficiency.
The field-to-voltage conversion gain of the link is then
\begin{equation}
G_V \;=\; \Rpd\,\Pdet\,\kappa_T\,Z_T
\qquad [\mathrm{V/(V/m)}],
\label{eq:GV}
\end{equation}
and, as in a photonic link, the gain scales with the detected optical
power $\Pdet$.

Each figure of merit of the receiver is a property of $T$. The conversion
gain and the sensitivity follow from its first field derivative at the bias
point \eqref{eq:GV}. The instantaneous bandwidth and the group delay follow
from the linearized dynamics of the atomic coherence that produces $T$. The
noise floor comprises whatever fluctuates the detected power or mimics a
displacement of the transmission feature (photon shot noise, laser
intensity noise, and laser frequency noise), each referred to the field
through the same chain \eqref{eq:kTdef}. Compression, the intercept points,
and the SFDR follow from the higher field derivatives of $T$ about the
bias. A known dipole matrix element and two laser wavelengths fix the field
dependence of $T$, with no calibration against a reference field, which is
the origin of the SI traceability.

\subsection{Transmission function}
\label{sec:transmission}

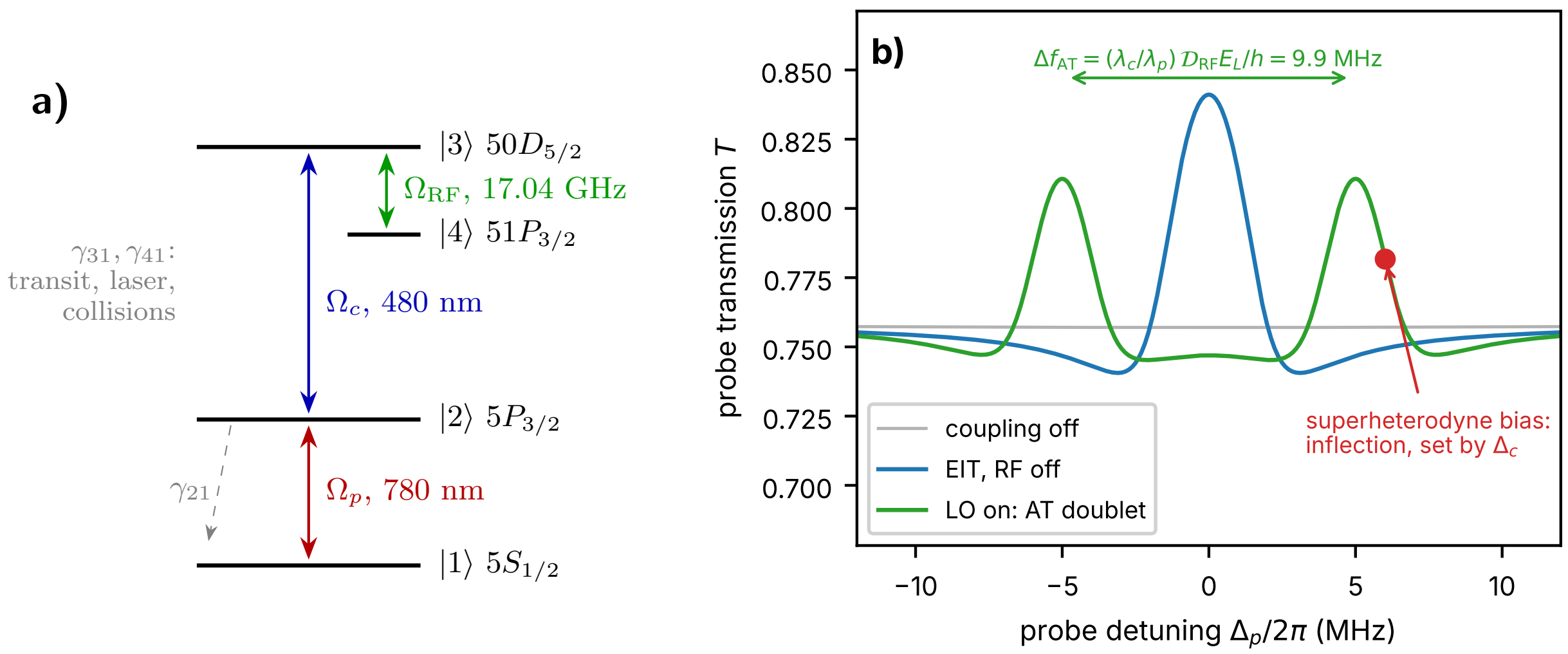
\begin{figure}[t]
\centering
\begin{minipage}[c]{0.42\textwidth}
\centering
\begin{tikzpicture}[
    lvl/.style={very thick},
    drv/.style={Stealth-Stealth, thick},
    lab/.style={font=\footnotesize},
]
\node[font=\large\bfseries\sffamily] at (-1.5, 4.75) {a)};
\draw[lvl] (0,0)   -- (2.3,0)   node[lab, right=1pt] {$|1\rangle\ 5S_{1/2}$};
\draw[lvl] (0,1.5) -- (2.3,1.5) node[lab, right=1pt] {$|2\rangle\ 5P_{3/2}$};
\draw[lvl] (0,4.3) -- (2.3,4.3) node[lab, right=1pt] {$|3\rangle\ 50D_{5/2}$};
\draw[lvl] (1.55,3.4) -- (2.3,3.4) node[lab, right=1pt] {$|4\rangle\ 51P_{3/2}$};
\draw[drv, red!70!black]  (1.15,0.06) -- (1.15,1.44)
    node[lab, midway, right=1pt, text=red!70!black] {$\Omega_p$, 780~nm};
\draw[drv, blue!70!black] (1.15,1.56) -- (1.15,4.24)
    node[lab, pos=0.42, right=1pt, text=blue!70!black] {$\Omega_c$, 480~nm};
\draw[drv, green!60!black] (1.95,3.46) -- (1.95,4.24)
    node[lab, midway, right=1pt, text=green!60!black]
    {$\Omega_{\mathrm{RF}}$, 17.04~GHz};
\draw[-Stealth, dashed, gray] (0.35,1.44) -- (0.12,0.25);
\node[lab, text=gray, anchor=east] at (0.30,0.75) {$\gamma_{21}$};
\node[lab, text=gray, align=right, anchor=east] at (-0.08,2.9)
    {\scriptsize $\gamma_{31},\gamma_{41}$:\\[-2pt]\scriptsize transit, laser,\\[-2pt]\scriptsize collisions};
\end{tikzpicture}
\end{minipage}\hfill
\begin{minipage}[c]{0.56\textwidth}
\centering
\includegraphics[width=\linewidth]{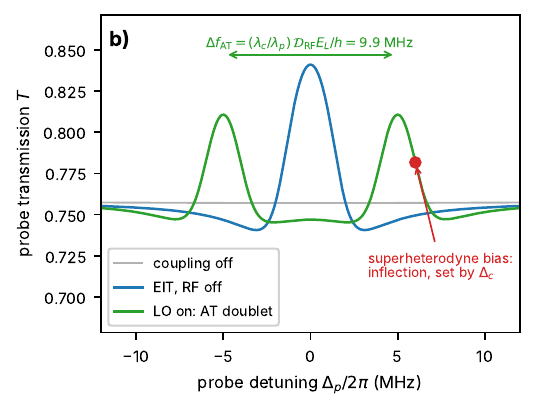}
\end{minipage}
\caption{The transducer and its operating point. (a) Four-level system of
the design example, with the optical and RF drives and the decay and
dephasing channels that parameterize
\eqref{eq:eom21}--\eqref{eq:eom41}. The $51P_{3/2}$ state lies 17.04~GHz
below $50D_{5/2}$. (b) Probe transmission versus
probe detuning at the design point, averaged over the
Maxwell--Boltzmann distribution [\eqref{eq:rho21cf}--\eqref{eq:T} with
\eqref{eq:doppleravg}, for $\gamma_{21}/2\pi=3$~MHz,
$\gamma_{31}/2\pi=\gamma_{41}/2\pi=0.344$~MHz, $\Omega_c/2\pi=3.5$~MHz,
and $d_0=0.28$]. The curves show the coupling off (gray), the EIT window
with the RF off (blue), and the Autler--Townes doublet with the 0.8-V/m LO
on (green), whose probe-scan separation is
$\Delta f_{\mathrm{AT}}=(\lambda_c/\lambda_p)\dip_{\mathrm{RF}}E_L/h
=9.9$~MHz \eqref{eq:ATsplit}. Superheterodyne operation
sets the two-photon detuning, via the coupling detuning $\Delta_c$, at the
inflection of one dressed feature (marker, drawn at the equivalent probe
detuning $(\lambda_c/\lambda_p)\Delta_c$), where the discriminator slope
$\kappa_T$ is maximal.}
\label{fig:ladder}
\end{figure}

The medium that sets $T$ is a four-level system in an alkali vapor
[Fig.~\ref{fig:ladder}(a)]. A weak probe couples the ground state
$|1\rangle$ to a short-lived intermediate state $|2\rangle$. A strong
coupling laser couples $|2\rangle$ to a Rydberg state $|3\rangle$. The
microwave field to be measured couples $|3\rangle$ to a neighboring Rydberg
state $|4\rangle$, which for the pair used below lies 17.04~GHz beneath it.
The choice of principal quantum number fixes jointly the RF carrier
($f_{\mathrm{RF}}\propto n^{*-3}$) and the coupling strength
($\dip\propto n^{*2}$). The scaling laws, the angular content of the dipole
moment, and the coherence-decay rates $\gamma_{21},\gamma_{31},\gamma_{41}$
that appear below are discussed in Appendix~\ref{app:atomic}. The four-level
system is treated once, and the bare EIT (three-level) system is recovered
as its RF-off limit.

The atomic state is a Hermitian matrix $\rho$ with $\mathrm{tr}\,\rho=1$,
whose diagonal entries are populations and whose off-diagonal entries are
coherences. A laser on the $i$--$j$ transition experiences a susceptibility
proportional to $\rho_{ij}$. The dynamics
$\dot\rho = -\tfrac{i}{\hbar}[H,\rho]+\mathcal{L}[\rho]$ form a linear
state-space model in which the commutator is the energy-conserving coupling
and the Lindblad term contributes linear damping,
$\mathcal{L}[\rho]_{ij}=-\gamma_{ij}\rho_{ij}$. In the rotating frame and
rotating-wave approximation, with the dipole coupling $-\mathbf{d}\cdot\mathbf{E}$, the
Hamiltonian has the nonzero elements
$H_{12}=H_{21}=-\hbar\Omega_p/2$, $H_{23}=H_{32}=-\hbar\Omega_c/2$, $H_{34}=H_{43}=-\hbar\ORF/2$,
$H_{22}=\hbar\Delta_p$, $H_{33}=\hbar(\Delta_p+\Delta_c)$, and
$H_{44}=\hbar(\Delta_p+\Delta_c+\dRF)$. The optical detunings are those of
the atomic transitions from the lasers, $\Delta_p=\omega_{21}-\omega_p$ and
$\Delta_c=\omega_{32}-\omega_c$. The RF detuning is
$\dRF=\omega_{\mathrm{RF}}-\omega_{43}$, where $\omega_{43}>0$ is the
frequency of the $|3\rangle$--$|4\rangle$ transition, whose lower state is
$|4\rangle$.

Evaluating $[H,\rho]_{ab}=\sum_k (H_{ak}\rho_{kb}-\rho_{ak}H_{kb})$ and
adding the damping via
$\dot\rho_{ab}=-\tfrac{i}{\hbar}[H,\rho]_{ab}+\mathcal{L}[\rho]_{ab}$
gives, to first order in the probe field, the equations of motion of the
probe, two-photon, and RF coherences,
\begin{align}
\dot\rho_{21} &= -(\gamma_{21}+i\Delta_p)\rho_{21}
+\tfrac{i}{2}\Omega_p(\rho_{11}-\rho_{22})
+\tfrac{i}{2}\Omega_c\rho_{31},
\label{eq:eom21}\\
\dot\rho_{31} &= -(\gamma_{31}+i\delta_2)\rho_{31}
+\tfrac{i}{2}\Omega_c\rho_{21}
+\tfrac{i}{2}\ORF\rho_{41},
\label{eq:eom31}\\
\dot\rho_{41} &= -(\gamma_{41}+i\delta_3)\rho_{41}
+\tfrac{i}{2}\ORF\rho_{31},
\label{eq:eom41}
\end{align}
with one-photon detuning $\Delta_p$, two-photon detuning
$\delta_2=\Delta_p+\Delta_c$, and three-photon detuning
$\delta_3=\delta_2+\dRF$. (The terms $-\tfrac{i}{2}\Omega_p\rho_{32}$ in
\eqref{eq:eom31} and $-\tfrac{i}{2}\Omega_p\rho_{42}$ in \eqref{eq:eom41}
are of second order in the probe and are omitted.)

The steady state of this chain telescopes into a closed form for the probe
coherence. From \eqref{eq:eom41},
$\rho_{41}=\frac{i\ORF/2}{\gamma_{41}+i\delta_3}\,\rho_{31}$. Substituting
into \eqref{eq:eom31} dresses the decay of the Rydberg coherence,
\begin{equation}
\gamma_{31}+i\delta_2 \;\longrightarrow\;
\gamma_{31}+i\delta_2+\frac{\ORF^2/4}{\gamma_{41}+i\delta_3},
\label{eq:dressdecay}
\end{equation}
and the steady state of \eqref{eq:eom31} then gives
\begin{equation}
\rho_{31}
= \frac{i\Omega_c/2}
{\gamma_{31}+i\delta_2+\dfrac{\ORF^2/4}{\gamma_{41}+i\delta_3}}\;\rho_{21}.
\label{eq:rho31}
\end{equation}
Inserting this into the steady state of \eqref{eq:eom21} in the weak-probe
limit ($\rho_{11}\approx 1$, $\rho_{22}\approx 0$),
\begin{equation}
(\gamma_{21}+i\Delta_p)\,\rho_{21}
= \tfrac{i}{2}\Omega_p
- \frac{\Omega_c^2/4}
{\gamma_{31}+i\delta_2+\dfrac{\ORF^2/4}{\gamma_{41}+i\delta_3}}\;\rho_{21},
\label{eq:collect}
\end{equation}
and collecting the $\rho_{21}$ terms yields the continued fraction
\begin{equation}
\rho_{21}
= \frac{i\Omega_p/2}
{(\gamma_{21}+i\Delta_p)
+ \dfrac{\Omega_c^2/4}
{\gamma_{31}+i\delta_2+\dfrac{\ORF^2/4}{\gamma_{41}+i\delta_3}}}.
\label{eq:rho21cf}
\end{equation}
Setting $\ORF\to 0$ removes the innermost term and recovers the three-level
EIT coherence.

The coherence becomes an observable through the macroscopic polarization. In
the half-amplitude convention
$E(t)=\tfrac12 E_p e^{-i\omega_p t}+\mathrm{c.c.}$, the polarization is
$P_p=\varepsilon_0\chi E_p$, and microscopically
$\tfrac12 P_p = \Nd\dip_{12}\rho_{21}$ for an atomic number density $\Nd$.
With $E_p=\hbar\Omega_p/\dip_{12}$,
\begin{equation}
\chi
= \frac{2\Nd\dip_{12}\rho_{21}}{\varepsilon_0 E_p}
= \frac{2\Nd\dip_{12}^2}{\varepsilon_0\hbar}\,\frac{\rho_{21}}{\Omega_p}
= \frac{i\Nd\dip_{12}^2}{\varepsilon_0\hbar\,\den},
\label{eq:chi}
\end{equation}
where $\den$ is the denominator of \eqref{eq:rho21cf}. The probe intensity
absorption coefficient is $\alpha=k_p\,\mathrm{Im}\,\chi$, and over a cell of
length $L$ the transmission of the cascade \eqref{eq:cascade} is
\begin{equation}
T \;=\; e^{-\alpha L} \;=\; \exp\!\big[-k_p L\,\mathrm{Im}\,\chi(\ERF,\nu)\big],
\label{eq:T}
\end{equation}
with the refractive phase governed by $\mathrm{Re}\,\chi$. Equations
\eqref{eq:rho21cf}--\eqref{eq:T} are the modulator transfer characteristic of
the link.

\subsection{The warm-vapor window}
\label{sec:contrast}

In a warm vapor the transfer characteristic is the average of
\eqref{eq:rho21cf} over the atomic velocities, and this average can be
carried out in closed form. For counter-propagating beams an atom of axial
velocity $v$ (measured along the coupling beam) sees the probe at
$\Delta_p-k_pv$ and all two-photon detunings
shifted by the residual $(k_c-k_p)v$, so the absorption is the
Maxwell--Boltzmann average
\begin{equation}
\bar\alpha(\Delta_p,\Delta_c)=\int\frac{e^{-v^2/u^2}}{u\sqrt{\pi}}\,
\mathrm{Re}\,\frac{\gamma_{21}}{\den(v)}\,\mathrm{d}v,
\label{eq:doppleravg}
\end{equation}
where $\den(v)$ is the denominator of \eqref{eq:rho21cf} with
$\Delta_p\to\Delta_p-k_pv$ and $\delta_{2,3}\to\delta_{2,3}+(k_c-k_p)v$,
$u=\sqrt{2k_BT_{\mathrm{cell}}/m}$ is the most probable speed at the cell
temperature $T_{\mathrm{cell}}$, and $\bar\alpha$ is expressed in units of
the resonant absorption coefficient of a single velocity class,
$\Nd\dip_{12}^2k_p/\varepsilon_0\hbar\gamma_{21}$. The homogeneous
structure is a
few megahertz wide, against a Doppler width of hundreds of megahertz
($k_pu/2\pi=303$~MHz for Rb at 293~K), so the Maxwell--Boltzmann weight is
constant across it. Changing variable to $x=k_pv-\Delta_p$, the two-photon
detuning of each class becomes $\xi+r x$, with
\begin{equation}
\xi\equiv\frac{\lambda_p}{\lambda_c}\,\Delta_p+\Delta_c,
\qquad
r\equiv\frac{\lambda_p}{\lambda_c}-1,
\label{eq:xi}
\end{equation}
so the transmission depends on the two laser detunings only through $\xi$.
The remaining integral over $x$ is of a rational function and follows by
residues (Appendix~\ref{app:numerics}). With the RF off it gives
\begin{equation}
\bar A(\xi) = \mathrm{Re}\,\frac{Z}{\sqrt{Z^2+K^2}},
\qquad
Z = r\gamma_{21}+\gamma_{31}+i\xi,
\qquad
K^2 = r\,\Omega_c^2,
\label{eq:Aflat}
\end{equation}
where $\bar A\equiv\bar\alpha/\bar\alpha_0$ is the absorption normalized to
its line-center value $\bar\alpha_0$ with the coupling off, so that
$T=e^{-d_0\bar A}$, and the root is taken in the right half-plane. The
wavelength mismatch ($r=0.625$ for the 780/480-nm ladder) enters twice, as
the additive width $r\gamma_{21}$ and as the factor $r$ on the coupling.
Equation~\eqref{eq:Aflat} matches the full velocity average to
$4\times10^{-3}$ in $\bar A$ at the design point (Appendix~\ref{app:numerics}).

The width of the window sets the responsivity, the bandwidth, and the field
scale of the nonlinearity, and its depth sets the modulation contrast
available to be read out. Because the transmission
depends on the detunings only through $\xi$, a coupling scan at fixed probe
and a probe scan at fixed coupling trace the same curve, with the
probe-detuning axis compressed by $\lambda_c/\lambda_p$. A window of full
width at half maximum $W_\xi$ in $\xi$ appears with
$W_p=(\lambda_c/\lambda_p)W_\xi$ in a probe scan. Widths and splittings are
compressed together, so their ratio does not depend on the scan. However,
the two widths play different roles below. $W_p$ sets the static field
scales, which are referred to splittings measured in a probe scan, while
$W_\xi$ sets the dynamics, because a field modulation moves every velocity
class through $\xi$. At the design point of Sec.~\ref{sec:worked},
$W_\xi=4.11$~MHz and $W_p=2.53$~MHz.

The depth follows from \eqref{eq:Aflat} at $\xi=0$,
\begin{equation}
1-\bar A(0)=1-\frac{1}{\sqrt{1+\mathcal{C}_D}},
\qquad
\mathcal{C}_D\equiv\frac{K^2}{(\mathrm{Re}\,Z)^2}
=\frac{r\,\Omega_c^2}{(r\gamma_{21}+\gamma_{31})^2},
\label{eq:CD}
\end{equation}
with $\mathcal{C}_D$ a Doppler cooperativity, the warm-vapor counterpart of
the cooperativity $\Omega_c^2/4\gamma_{21}\gamma_{31}$ of a single velocity
class (the ratio of the coherent two-photon coupling to the damping it
bridges). Far from two-photon resonance the coupling term vanishes and the
absorption returns to its bare value, which defines the background optical
depth
\begin{equation}
d_0 \equiv -\ln T\,\Big|_{\Omega_c\to0,\ \Delta_p=0}
= \Nd\sigma_0 L,
\qquad
\sigma_0=\frac{3\lambda_p^2}{2\pi}\,R.
\label{eq:d0}
\end{equation}
A single velocity class of a closed transition has the resonant
cross-section $\dip_{12}^2k_p/\varepsilon_0\hbar\gamma_{21}=3\lambda_p^2/2\pi$.
The Doppler average reduces it by
$\bar\alpha_0\approx\sqrt{\pi}\,\gamma_{21}/k_pu\approx1.7\times10^{-2}$, and
hyperfine and transition-strength branching reduce it further. The factor
$R\le1$ collects these reductions, and $R\approx10^{-2}$ is taken below.
With transparency peak $T_{\mathrm{peak}}=e^{-d_0\bar A(0)}$ on the
background $T_b=e^{-d_0}$, the contrast is
\begin{equation}
\CEIT = \frac{T_{\mathrm{peak}}-T_b}{T_{\mathrm{peak}}}
= 1-\exp\!\Big\{-d_0\Big[1-\frac{1}{\sqrt{1+\mathcal{C}_D}}\Big]\Big\}.
\label{eq:contrast}
\end{equation}
The contrast grows linearly with the optical depth while it is small and
rises with the coupling through $\mathcal{C}_D$, saturating at
$1-e^{-d_0}$ only for $\mathcal{C}_D\gg1$. At the design point
($\gamma_{31}/2\pi=0.344$~MHz, Sec.~\ref{sec:worked}) $\mathcal{C}_D=1.55$,
the depth $1-\bar A(0)$ is 0.374, and a contrast $\CEIT=0.1$ requires
$d_0=0.28$. This optical depth is reached for a number density
$\Nd\approx2\times10^{15}~\mathrm{m^{-3}}$
($\approx2\times10^{9}~\mathrm{cm^{-3}}$) over a
path $L\approx5$~cm, for which a single resonant velocity class would have
an optical depth of 28. These are ordinary warm-cell conditions, so
$\CEIT=0.1$ is a conservative choice. Raising $d_0$
toward unity would raise $\CEIT$ toward a few tenths, with the responsivity
of Sec.~\ref{sec:responsivity} growing in proportion.

\subsection{Field map and local oscillator}
\label{sec:fieldmap}\label{sec:superhet}

The RF field enters
the cascade \eqref{eq:cascade} by displacing the transmission feature in the
frequency coordinate. The responsivity defined in
\eqref{eq:kTdef} factors by the chain rule into a lineshape
(discriminator) slope and a field-to-frequency (displacement) slope,
\begin{equation}
\kappa_T = \frac{1}{T}\frac{\partial T}{\partial \ERF}
= \underbrace{\Big(\frac{1}{T}\frac{\partial T}{\partial\nu}\Big)}_{\text{discriminator slope}}
\;\underbrace{\frac{\partial\nu}{\partial \ERF}}_{\text{displacement slope}}.
\label{eq:chainrule}
\end{equation}
Both factors follow in closed form.

The displacement slope is a property of the RF-dressed Rydberg pair. The RF
field couples
$|3\rangle$ and $|4\rangle$ with Rabi frequency
$\ORF=\dip_{\mathrm{RF}}\ERF/\hbar$ and detuning $\dRF=\omega_{\mathrm{RF}}-\omega_{43}$
(the dipole $\dip_{\mathrm{RF}}$ is computed in Appendix~\ref{app:dipole}).
In the RF rotating frame, the two-state Hamiltonian for
$\{|3\rangle,|4\rangle\}$, referred to $|3\rangle$, is
\begin{equation}
H_{34} = \hbar\begin{pmatrix} 0 & -\ORF/2 \\ -\ORF/2 & \dRF \end{pmatrix}.
\label{eq:H34}
\end{equation}
Its eigenvalues solve $\det(H_{34}-\En\,\mathbf{1})=0$, which is
$(\En/\hbar)^2-\dRF\,(\En/\hbar)-\ORF^2/4=0$, giving
\begin{equation}
\frac{\En_\pm}{\hbar}
= \frac12\Big(\dRF\pm\sqrt{\dRF^2+\ORF^2}\Big)
\equiv \frac12\big(\dRF\pm\tilde\Omega\big),
\label{eq:Epm}
\end{equation}
with the generalized Rabi frequency $\tilde\Omega=\sqrt{\dRF^2+\ORF^2}$. The
eigenvectors mix $|3\rangle$ and $|4\rangle$ through the angle $\theta_d$,
\begin{equation}
\tan\theta_d = \frac{\ORF}{\dRF},\qquad
\sin\theta_d = \frac{\ORF}{\tilde\Omega},\qquad
\cos\theta_d = \frac{\dRF}{\tilde\Omega},
\label{eq:thetad}
\end{equation}
and the $|3\rangle$-content of the two dressed states is $\cos^2(\theta_d/2)$
and $\sin^2(\theta_d/2)$. Because the coupling laser reaches the Rydberg
manifold through $|3\rangle$, this content sets the relative heights of the
two EIT features and with them the feature that carries the steeper
readable slope.

The dressed energies become observable feature positions once the optical
readout geometry is imposed. EIT occurs where the two-photon detuning cancels
a dressed Rydberg energy. The averaged transmission depends on the detunings
only through $\xi$ \eqref{eq:xi}, and its dressed features appear at
$\xi=-\En_\pm/\hbar$, so with the coupling on resonance they sit at probe
detuning $\Delta_p=-(\lambda_c/\lambda_p)\En_\pm/\hbar$. Physically, the
probe selects the velocity class that is resonant with the
one-photon transition, $k_pv=\Delta_p$, and the residual two-photon Doppler
shift of that
class compresses every splitting by $\lambda_c/\lambda_p$. Converting to
ordinary frequency and writing the Rabi frequency and detuning in hertz,
$f_R\equiv\ORF/2\pi=\dip_{\mathrm{RF}}\ERF/h$ and
$\delta_{\mathrm{RF}}\equiv\dRF/2\pi$, the two feature positions are
\begin{equation}
\nu^{(\pm)} = \frac{\lambda_c}{\lambda_p}\,
\frac12\Big(-\delta_{\mathrm{RF}}\pm\sqrt{\delta_{\mathrm{RF}}^2+f_R^2}\Big),
\qquad
f_R=\frac{\dip_{\mathrm{RF}}}{h}\ERF.
\label{eq:nupm}
\end{equation}
On RF resonance ($\delta_{\mathrm{RF}}=0$) the peaks are at
$\nu^{(\pm)}=\pm\tfrac12(\lambda_c/\lambda_p)f_R$ and their separation is the
Autler--Townes splitting
\begin{equation}
\Delta f_{\mathrm{AT}} = \nu^{(+)}-\nu^{(-)}
= \frac{\lambda_c}{\lambda_p}\,\frac{\dip_{\mathrm{RF}}\ERF}{h},
\label{eq:ATsplit}
\end{equation}
which inverts to give the field in SI units,
\begin{equation}
\ERF = \frac{h}{\dip_{\mathrm{RF}}}\,\frac{\lambda_p}{\lambda_c}\,\Delta f_{\mathrm{AT}}.
\label{eq:Einv}
\end{equation}
A measured frequency splitting and two known optical wavelengths determine the
field strength without reference to a calibrated RF source. The
small-signal slope of the splitting,
\begin{equation}
\kappa_\nu \equiv \frac{\partial(\Delta f_{\mathrm{AT}})}{\partial \ERF}
= \frac{\lambda_c}{\lambda_p}\,\frac{\dip_{\mathrm{RF}}}{h}
\qquad [\mathrm{Hz/(V/m)}],
\label{eq:kappanu}
\end{equation}
is the AT responsivity. For a single-velocity (cold) sample the compression
is absent and $\kappa_\nu\to\dip_{\mathrm{RF}}/h$.

Metrological readout of the splitting and small-signal reception place
different demands on this map. A weak field does not resolve the doublet,
and the line-center transmission is then an even function of $\ERF$ whose
first derivative vanishes at zero field, so a direct readout responds to a
weak signal only at second order. As in a coherent
receiver, this is remedied by supplying a strong, phase-coherent LO at the
RF carrier and recovering the beat between the LO and
signal fields \cite{simons_rydberg_2019}. For collinear, co-polarized fields
with $E_S\ll E_L$, the
instantaneous amplitude is
\begin{equation}
\begin{split}
\ERF(t) &= \big|E_L+E_S e^{i\Omega t}\big|
= E_L\sqrt{1+2\frac{E_S}{E_L}\cos\Omega t+\frac{E_S^2}{E_L^2}}\\
&= E_L+E_S\cos\Omega t+\mathcal{O}\Big(\frac{E_S^2}{E_L}\Big),
\end{split}
\label{eq:beat}
\end{equation}
a tone at the offset (IF) frequency $\Omega$ riding on the bias $E_L$
\cite{simons_rydberg_2019,jing_atomic_2020}. The $\mathcal{O}(E_S^2/E_L)$
term returns in
Sec.~\ref{sec:nonlin} as the dominant source of even-order distortion. This
$\Omega$ is the modulation frequency that the transfer function
$H(\Omega)$ of Sec.~\ref{sec:dynamics} rolls off. The observable is the
displacement of
the single dressed feature on which the probe sits, and the peak-to-peak
splitting is not read. Differentiating \eqref{eq:nupm} for the upper feature,
\begin{equation}
\frac{\partial\nu^{(+)}}{\partial \ERF}
= \frac{\lambda_c}{\lambda_p}\,\frac12\,
\frac{\partial}{\partial \ERF}\sqrt{\delta_{\mathrm{RF}}^2+f_R^2}
= \frac{\lambda_c}{\lambda_p}\,\frac12\,
\frac{f_R}{\sqrt{\delta_{\mathrm{RF}}^2+f_R^2}}\,\frac{\dip_{\mathrm{RF}}}{h},
\label{eq:dnudE}
\end{equation}
and evaluating at the LO operating point, where
$f_R=\dip_{\mathrm{RF}}E_L/h$ and the ratio in \eqref{eq:dnudE} is
$\sin\theta_d$ by \eqref{eq:thetad},
\begin{equation}
\frac{\partial\nu}{\partial \ERF}\bigg|_{\mathrm{single}}
= \frac12\sin\theta_d\,\kappa_\nu,
\label{eq:singleslope}
\end{equation}
where the factor $\tfrac12$ is the geometry of reading a single dressed
feature in place of the full splitting, and $\sin\theta_d\le1$ is the
dressed-state mixing set by the RF--LO detuning. The two features move
oppositely, $\partial\nu^{(\pm)}/\partial\ERF=\pm\tfrac12\sin\theta_d\,\kappa_\nu$
(the common-mode term $-\delta_{\mathrm{RF}}$ in \eqref{eq:nupm} is
field-independent). A differential readout with two probe beams, one on the
outer flank of one feature and one on the inner flank of the other,
subtracted, collects
\begin{equation}
\frac{\partial\nu}{\partial \ERF}\bigg|_{\mathrm{diff}}
= \sin\theta_d\,\kappa_\nu,
\label{eq:balanced}
\end{equation}
while the intensity and frequency noise common to the two beams cancel. The
beams carry independent shot noise, so this readout lowers the shot-limited
floor by $\sqrt2$ at twice the detected power. The budget of
Sec.~\ref{sec:worked} uses the single-feature readout and quotes the
differential values in parentheses.

The LO amplitude is a design variable, and, like the LO drive of a mixer, it
sets the linearity of the conversion. The single-feature expressions
above hold once the doublet is resolved. The numerical model of
Appendix~\ref{app:numerics} gives an outer-flank slope within 1\% of its
resolved value for $E_L\gtrsim0.26$~V/m ($\ORF/2\pi\gtrsim1.25\,W_\xi$)
and constant to 0.2\% above 0.4~V/m. On this plateau the slope no longer
depends on $E_L$, while the second-order intercept grows with it
(Sec.~\ref{sec:nonlin}), so a single-feature bias favors a strong LO. The
design adopts
\begin{equation}
E_L = 0.8~\mathrm{V/m},
\label{eq:ELrule}
\end{equation}
well inside the plateau. The alternative holds both lasers on resonance and
lets a weak LO split the window symmetrically, as in the superheterodyne
demonstration of Ref.~\cite{jing_atomic_2020}. The slope of that splitting
peaks at a probe-scan splitting $\Delta f_{\mathrm{AT}}\approx0.8\,W_p$
($E_L^\star\approx0.16$~V/m in the weak-probe limit and 0.23~V/m at the
operating probe strength, for which the probe broadens the window to
$W_p=3.5$~MHz), where it is 1.5--1.6 times the single-feature
slope. While this line-center bias also rejects laser frequency noise to
first order (Sec.~\ref{sec:noise}) and needs a local oscillator 11--14~dB weaker,
its second-order intercept is only about $2E_L^\star$. The closed
forms below
refer to the single-feature bias, and the line-center values quoted are
those of the numerical model (Appendix~\ref{app:numerics}).

\subsection{Responsivity}
\label{sec:responsivity}

The remaining factor of the chain rule \eqref{eq:chainrule} is the
logarithmic
slope of the transparency feature at the bias. For a feature of fractional
height $\CEIT$ and full width $W$ (in hertz) it is convenient to write its
maximum as
\begin{equation}
\max_\nu\bigg|\frac{1}{T}\frac{\partial T}{\partial\nu}\bigg|
= \kappa_s\,\frac{\CEIT}{W},
\label{eq:kappasdef}
\end{equation}
with $\kappa_s$ a dimensionless lineshape constant. For a Lorentzian
transparency feature of small height on the absorbing background
$T_b=e^{-d_0}$, to first order in $\CEIT$,
\begin{equation}
T(\nu) \simeq T_b\big[1+\CEIT\,L(\nu)\big],
\qquad
L(\nu)=\frac{1}{1+[2(\nu-\nu_0)/W]^2},
\label{eq:lorentz}
\end{equation}
differentiating with $y\equiv 2(\nu-\nu_0)/W$ gives
\begin{equation}
\frac{1}{T}\frac{\partial T}{\partial\nu}
= \frac{\CEIT\,L'(\nu)}{1+\CEIT L(\nu)},
\qquad
L'(\nu) = \frac{-2y}{(1+y^2)^2}\cdot\frac{2}{W}.
\label{eq:dTdnu}
\end{equation}
For the small-to-moderate contrast of interest ($\CEIT\lesssim0.3$) the
denominator is $1+\mathcal{O}(\CEIT)$ and the extremum of the slope is set by
$L'$. Setting $\mathrm{d}[2y/(1+y^2)^2]/\mathrm{d}y=0$ gives
$1-3y^2=0$, so the inflection sits at $y=1/\sqrt{3}$, where
$|{-2y}/(1+y^2)^2|=(2/\sqrt{3})/(4/3)^2=9/8\sqrt{3}=0.6495$. Hence
\begin{equation}
\kappa_s = 2\cdot\frac{9}{8\sqrt{3}} = \frac{9}{4\sqrt{3}} \simeq 1.30
\qquad(\text{Lorentzian}).
\label{eq:kappas}
\end{equation}
For a dispersive (phase) discriminator or a derivative (lock-in) readout the
constant differs but remains $\mathcal{O}(1)$.

The dressed feature of the warm vapor is not Lorentzian, but its slope is
also available in closed form. For a resolved doublet with a resonant LO and
$\gamma_{41}=\gamma_{31}$, the dressing shares the coupling equally between
the two dressed states, and each feature has the form \eqref{eq:Aflat} with
\begin{equation}
Z\to Z_\pm = r\gamma_{21}+\frac{\gamma_{31}+\gamma_{41}}{2}
+i\Big(\xi\mp\frac{\ORF}{2}\Big),
\qquad
K^2\to K_d^2=\frac{K^2}{2}.
\label{eq:dressedflat}
\end{equation}
Differentiating with respect to $\xi$ and multiplying by the displacement of
the feature, $\partial\xi_+/\partial\ERF=\dip_{\mathrm{RF}}/2\hbar$ (the
two-photon form of \eqref{eq:singleslope}), gives the slope at the outer
inflection,
\begin{equation}
\kappa_T = \frac{d_0}{2}\,\frac{\dip_{\mathrm{RF}}}{\hbar}\,
\frac{\Phi(K_d/\Gamma_d)}{\Gamma_d},
\qquad
\Phi(k)\equiv\max_s\Big|\,k^2\,\mathrm{Im}\big[(1+is)^2+k^2\big]^{-3/2}\Big|,
\label{eq:kTflat}
\end{equation}
with $\Gamma_d=\mathrm{Re}\,Z_\pm$. The shape function $\Phi$ grows as
$0.73\,k^2$ for weak coupling and roughly as $0.31\sqrt{k}$ for strong
coupling, without a maximum. Written in the form \eqref{eq:kappasdef} with
the width and contrast of the RF-off window, \eqref{eq:kTflat} corresponds
to the lineshape constant
\begin{equation}
\kappa_s = \frac{d_0\,\Phi(K_d/\Gamma_d)}{\CEIT}\,\frac{2\pi W_\xi}{\Gamma_d},
\label{eq:kappasD}
\end{equation}
with $\Gamma_d$ an angular rate. At the design point $\kappa_s=1.47$. It
exceeds the Lorentzian value because the Doppler-averaged window has steeper
flanks than a Lorentzian of the same width ($\kappa_s=1.97$ for the RF-off
window itself), while sharing the coupling between the two dressed states,
$K_d^2=K^2/2$, makes each feature narrower but shallower and lowers the
constant. Over $\Omega_c/2\pi=1.5$--7~MHz and $\gamma_{31}/2\pi=0.3$--1.5~MHz it
ranges from 0.8 to 3.0.

With both factors in closed form, the responsivity assembles directly.
Substituting \eqref{eq:kappasdef} with $W=W_p$, the width in the coordinate
in which $\kappa_\nu$ is defined, and \eqref{eq:singleslope} into the chain
rule \eqref{eq:chainrule} gives the superheterodyne responsivity, biased at
the inflection of the chosen dressed feature [Fig.~\ref{fig:ladder}(b)],
\begin{equation}
\kappa_T = \kappa_s\,\frac{\CEIT}{W_p}\cdot\frac{\sin\theta_d}{2}\,\kappa_\nu
\qquad [(\mathrm{V/m})^{-1}],
\label{eq:kT}
\end{equation}
with $\kappa_s$ from \eqref{eq:kappasD}, and with
$\sin\theta_d/2\to\sin\theta_d$ for the differential readout \eqref{eq:balanced}.
Equation~\eqref{eq:kT} is the modulation slope of the link and, inserted into
\eqref{eq:GV}, it completes the small-signal cascade from field to output
voltage.

At fixed optical depth the slope
rises monotonically with the coupling strength, because in \eqref{eq:kTflat}
$\Gamma_d$ does not depend on $\Omega_c$, $K_d\propto\Omega_c$, and $\Phi$ has
no maximum. In practice the available coupling-laser power sets the
coupling Rabi frequency (optimal laser parameters for
the EIT readout are analyzed in Ref.~\cite{meyer_optimal_2021}). However,
\eqref{eq:kTflat} is a weak-probe slope. At the probe strength that minimizes
the shot-noise floor, $\Omega_p\approx0.9\,\Omega_c$, the Lindblad master
equation of Appendix~\ref{app:numerics} gives about half of it, and
the operating slope is written $\mathcal{S}\kappa_T$ with
$\mathcal{S}\approx\tfrac12$.

\subsection{Dynamics}
\label{sec:dynamics}

The electronics of the cascade \eqref{eq:cascade} are broadband, so
the speed of the receiver is the rate at which $T$ itself can change, which
is the rate at which the atomic coherence follows the field. The mechanism
is clearest for a single velocity class. In the superheterodyne regime the RF
field displaces the EIT resonance (it shifts the two-photon detuning in
proportion to the field), $\delta(t)=\delta_b+s\,\delta E(t)$, where
$s=\dip_{\mathrm{RF}}\sin\theta_d/2\hbar$ is the displacement slope
\eqref{eq:singleslope} in the two-photon coordinate and
$\delta E(t)=\delta E\,e^{i\Omega t}$ is the small
signal about the bias $\delta_b$. Writing $\rho=\rho^{(0)}+\delta\rho$ in
\eqref{eq:eom21}--\eqref{eq:eom31} and keeping first order, the static probe
drive $\Omega_p$ drops out, and (because $\delta$ multiplies
$\rho_{31}$) the field modulation appears as a source on the two-photon
coherence,
\begin{align}
\delta\dot\rho_{21} &= -(\gamma_{21}+i\Delta_p)\,\delta\rho_{21}
+\tfrac{i}{2}\Omega_c\,\delta\rho_{31},
\label{eq:lin21}\\
\delta\dot\rho_{31} &= -(\gamma_{31}+i\delta_b)\,\delta\rho_{31}
+\tfrac{i}{2}\Omega_c\,\delta\rho_{21}
\;\underbrace{-\,i\rho^{(0)}_{31}\,s\,\delta E(t)}_{\text{field drive}}.
\label{eq:lin31}
\end{align}
For a modulation $\propto e^{i\Omega t}$ the first equation gives
$\delta\rho_{21}=\frac{i\Omega_c/2}{\gamma_{21}+i\Delta_p+i\Omega}\,\delta\rho_{31}$,
and substituting into the second collapses the pair onto the two-photon
coherence,
\begin{equation}
\Big[\,i\Omega+\gamma_{31}+i\delta_b
+\underbrace{\frac{\Omega_c^2/4}{\gamma_{21}+i\Delta_p+i\Omega}}_{\text{optical coherence}}\Big]
\delta\rho_{31} = -i\rho^{(0)}_{31}\,s\,\delta E.
\label{eq:linfull}
\end{equation}
The transmission follows the optical coherence, so on the operating bias
the transfer function $\delta T(\Omega)=\kappa_T\,H(\Omega)\,\delta E$ is the
reciprocal of this bracket times $\gamma_{21}+i\Delta_p+i\Omega$, normalized to
its dc value. The bias detuning $\delta_b$ shifts the slow pole along the
imaginary axis, which makes the upper and lower sidebands of the LO see
different bandwidths (Appendix~\ref{app:numerics}). Setting it aside, with
$\Delta_p=0$, the product is a quadratic in $i\Omega$, and $H$ is a two-pole
response,
\begin{equation}
H(\Omega) = \frac{\lambda_-\lambda_+}{(\lambda_-+i\Omega)(\lambda_++i\Omega)},
\label{eq:Hexact}
\end{equation}
whose poles solve $(i\Omega+\gamma_{31})(i\Omega+\gamma_{21})+\tfrac14\Omega_c^2=0$,
which under $i\Omega\to-\lambda$ reads
\begin{equation}
\lambda^2-(\gamma_{21}+\gamma_{31})\lambda
+\Big(\gamma_{21}\gamma_{31}+\frac{\Omega_c^2}{4}\Big)=0,
\qquad
\lambda_\pm
= \frac{(\gamma_{21}+\gamma_{31})\pm\sqrt{(\gamma_{21}-\gamma_{31})^2-\Omega_c^2}}{2}.
\label{eq:poles}
\end{equation}
These are the decay rates of the two normal modes (bright and dark
polaritons) of the driven three-level system, the same roots that give the
absorption line of the class its shape. For $\Omega_c<\gamma_{21}-\gamma_{31}$
they are real, with a fast mode $\lambda_+\approx\gamma_{21}$ and a slow mode
which, for $\Omega_c\ll\gamma_{21}-\gamma_{31}$ and
$\gamma_{31}\ll\gamma_{21}$, is
\begin{equation}
\lambda_- \simeq \gamma_c\equiv\gamma_{31}+\frac{\Omega_c^2}{4\gamma_{21}}=\frac{\GEIT}{2},
\qquad
\GEIT = \frac{\Omega_c^2}{2\gamma_{21}}+2\gamma_{31}
= \frac{\Omega_c^2}{\Gamma_e}+2\gamma_{31},
\label{eq:GammaEIT}
\end{equation}
the width of the EIT line of one velocity class
\cite{fleischhauer_electromagnetically_2005,gea-banacloche_electromagnetically_1995},
with $\Gamma_e\equiv2\gamma_{21}$ the natural width of the intermediate state.
(Beyond $\Omega_c=\gamma_{21}-\gamma_{31}$ the poles are complex and the line
of the class splits into an Autler--Townes doublet \cite{autler_stark_1955}.)
At weak coupling the slow mode dominates the response,
\begin{equation}
H(\Omega) \;\simeq\; \frac{1}{1+i\Omega/\gamma_c},
\qquad
\tau_g(0) = \frac{1}{\lambda_-}+\frac{1}{\lambda_+}
\simeq \frac{1}{\gamma_c},
\label{eq:Hsingle}
\end{equation}
the adiabatic limit in which the optical coherence follows the two-photon
coherence. The fast mode adds $1/\lambda_+\approx1/\gamma_{21}$ to the delay
and steepens the roll-off beyond $\gamma_{21}$.

In the vapor every velocity class is driven by the same modulation of $\xi$,
and the measured transmission is the average of the class responses. The
width of the averaged window in $\xi$ then takes the place of the linewidth of
a single class, and the single-pole form \eqref{eq:Hsingle} applies, to a good
approximation, with $\gamma_c\to\pi W_\xi$. The half-power frequency, where
$|H|^2=[1+(\Omega/\gamma_c)^2]^{-1}=\tfrac12$, is then
\begin{equation}
f_{3\mathrm{dB}} = \frac{\gamma_c}{2\pi} = \frac{W_\xi}{2},
\label{eq:f3dB}
\end{equation}
and the phase of $H$ gives the group delay
\begin{equation}
\tau_g(\Omega) = -\frac{\mathrm{d}\arg H}{\mathrm{d}\Omega}
= \frac{1/\gamma_c}{1+(\Omega/\gamma_c)^2},
\qquad
\tau_g(0)=\frac{1}{\pi W_\xi}.
\label{eq:taug}
\end{equation}
For a weak probe the numerical model of Appendix~\ref{app:numerics}
reproduces this bandwidth
to 10\% but gives a delay 50\% longer, a difference attributed there to the
superposition of velocity classes. The double-sided IF passband around the
LO then spans $\approx W_\xi$. However, the rule does not survive a strong
probe. At the operating probe strength of
Sec.~\ref{sec:worked} the probe broadens the window to $W_\xi=5.7$~MHz while
the bandwidth falls to 1.6~MHz, 0.57 of half the broadened width, a loss
that is attributed to saturation.

The delay limits the agility for fast frequency hopping and dechirping,
because the operating point cannot be retuned, nor the field tracked, faster
than $\sim1/\pi W_\xi$, and a large change of operating point settles more
slowly still (Sec.~\ref{sec:validation}). The instantaneous bandwidth must also not be confused
with the tuning coverage, which reaches from 1~GHz to 500~GHz in a single
cell by choice of $n$ \cite{holloway_broadband_2014}. The receiver is
wideband in carrier choice but narrowband instantaneously. A signal
broader than the window requires multiple Rydberg lines or a comb of LO
tones, since a single EIT resonance cannot capture it in
one shot. Receivers that go beyond a single window have reached bandwidths of 10
to 210~MHz, with a frequency-comb probe
\cite{artusio-glimpse_increased_2024}, six-wave mixing
\cite{yang_highly_2024}, engineered multi-dressed states
\cite{yan_multi-dressed-state_2026}, or a Stark-comb array of vapor cells
\cite{jiao_-demand_2026}, and single-window receivers are compared with
\eqref{eq:f3dB} in Sec.~\ref{sec:validation}.

\section{Sensitivity and dynamic range}
\label{sec:sensdr}

The usable range of the receiver lies between a noise floor and a
distortion ceiling. Both are properties of $T$ and of the optical carrier,
and both translate into the power-referred conventions and cascaded budgets
of receiver engineering. The expressions of this section are general, and
Sec.~\ref{sec:worked} evaluates them.

\subsection{Noise floor}
\label{sec:noise}

The cascade \eqref{eq:cascade} recovers a small modulation on a large
optical carrier by square-law detection, so the fluctuations of the carrier
set the sensitivity, as in an analog optical link. Each contribution
is expressed as a noise-equivalent field (NEF), the rms field of a tone
that gives unity signal-to-noise ratio in 1~Hz, by referring the
photocurrent noise to the field through the signal current
$i_{\mathrm{sig}}=\Rpd\Pdet\,\kappa_T\,\delta E$ of \eqref{eq:kTdef}.

The irreducible optical term is shot noise on the detected carrier. A
current of mean $i_{\mathrm{dc}}=\Rpd\Pdet$ carries the one-sided Schottky
density $S_i=2e\,i_{\mathrm{dc}}$, so
\begin{equation}
\NEF_{\mathrm{PSN}}
= \frac{\sqrt{2e\,\Rpd\Pdet}}{\Rpd\Pdet\,\kappa_T}
= \frac{1}{\kappa_T}\sqrt{\frac{2h\nu_p}{\eta\Pdet}},
\label{eq:NEFpsn}
\end{equation}
using $\Rpd=\eta e/h\nu_p$ from \eqref{eq:current}. It improves only as the
square root of the detected power, the scaling characteristic of an IMDD
link.

The relative intensity noise of the probe,
\begin{equation}
\RIN(f) \equiv \frac{S_{\Delta P}(f)}{\langle P\rangle^2}
\qquad [\mathrm{Hz^{-1}}],
\label{eq:RINdef}
\end{equation}
produces the photocurrent density $i_{\mathrm{dc}}^2\,\RIN$, which at the
detector is indistinguishable from a modulation of the transmission.
Referred through the same signal current, and evaluated at the IF offset
$\Omega$ to which the local oscillator translates the signal,
\begin{equation}
\NEF_{\RIN} = \frac{\sqrt{\RIN(\Omega)}}{\kappa_T}.
\label{eq:NEFrin}
\end{equation}
Unlike \eqref{eq:NEFpsn}, this floor does not improve with $\Pdet$. Raising
the carrier power lowers the shot-noise NEF only until the two terms cross,
at
\begin{equation}
P^\star_{\mathrm{det}} = \frac{2h\nu_p}{\eta\,\RIN(\Omega)},
\label{eq:Pstar}
\end{equation}
so the probe RIN is subdominant when it lies below the shot-noise-equivalent
value $\RIN_{\mathrm{eq}}=2h\nu_p/\eta\Pdet$. Two remedies familiar from
photonic links apply \cite{urick_fundamentals_2015,cox_analog_2004}. The IF
can be placed beyond the $1/f$ corner of the laser RIN and away from its
relaxation-oscillation peak, which is worth tens of decibels against
baseband operation and is, with the linear bias, a principal advantage of
heterodyne readout \cite{jing_atomic_2020}. Because RIN is multiplicative on
the carrier, it is also common-mode between the signal photodiode and a
reference tap of the same beam, so balanced detection or active intensity
stabilization suppresses it by 20--30~dB. Coupling-laser RIN enters at
first order through the depth and width of the window, with
$\mathrm{d}\ln T/\mathrm{d}\ln P_c=0.04$ for a weak probe and 0.02 at the
operating probe strength (Appendix~\ref{app:numerics}), 28 to 34~dB below
the unit coefficient of the probe. Amplitude noise on the local
oscillator enters by the same square-law route. Its AM sidebands at the IF
modulate the bias and contribute
$\NEF_{\mathrm{LO}}=E_L\sqrt{S_{\mathrm{AM}}(\Omega)}$
for a one-sided fractional amplitude-noise density $S_{\mathrm{AM}}$ (twice
the single-sideband level in dBc/Hz).

Frequency noise enters through the discriminator itself, which converts any
relative fluctuation between the lasers and the atomic feature into
apparent signal. The relevant fluctuation is the combination of the two
laser frequencies that enters $\xi$ \eqref{eq:xi}, referred to the probe
scan, $\delta\nu=\delta\nu_p+(\lambda_c/\lambda_p)\,\delta\nu_c$. It is
indistinguishable from a
field fluctuation $\delta\nu/(\partial\nu/\partial\ERF)$, so
\begin{equation}
\NEF_{\mathrm{FM}}
= \frac{\sqrt{S_\nu(\Omega)}}{\tfrac12\sin\theta_d\,\kappa_\nu},
\label{eq:NEFfm}
\end{equation}
with $S_\nu$ the one-sided density of $\delta\nu$ at the IF. The in-band
frequency noise of a free-running laser lies orders of magnitude above the
levels that matter here and would dominate the budget. For this reason the
superheterodyne measurements of Ref.~\cite{jing_atomic_2020} lock both
lasers to a common high-finesse cavity. Slow components of $S_\nu$ broaden
the window
instead. At the line-center bias of Sec.~\ref{sec:superhet} the
transmission is stationary with respect to both laser frequencies, and
\eqref{eq:NEFfm} vanishes to first order (Appendix~\ref{app:numerics}).

Collecting terms, the independent noises add in quadrature,
\begin{equation}
\NEF_{\mathrm{tot}}^2
= \frac{1}{\kappa_T^2}\Big[\frac{2h\nu_p}{\eta\Pdet}+\RIN(\Omega)\Big]
+\NEF_{\mathrm{FM}}^2,
\label{eq:NEFtot}
\end{equation}
to which detector dark current and the input noise of the transimpedance
stage add terms divided by $(\Rpd\Pdet)^2$. These are suppressed by a large
carrier power or, at low carrier power, by a high transimpedance
(Sec.~\ref{sec:worked}).

\subsection{Fundamental floors}
\label{sec:limits}

Beneath these technical floors lies the projection noise of the atomic
stage itself, the standard quantum limit (SQL) of an unentangled ensemble
\cite{itano_quantum_1993,degen_quantum_2017}. Each atom is read out as a
projective two-level measurement with outcome probability
$p(\phi)=\tfrac12(1+\cos\phi)$, where $\phi=2\pi\nu_sT_2$ is the phase
accumulated over the coherence time at the field-induced shift
$\nu_s=\dip_{\mathrm{RF}}\ERF/h$. The Fisher information per atom and shot,
$(\partial_\phi p)^2/p(1-p)=1$, is independent of $\phi$, so for $N$ atoms
and $T'/T_2$ shots the Cram\'er--Rao bound gives
\begin{equation}
\Delta\nu_{s,\min} = \frac{1}{2\pi\sqrt{N\,T_2\,T'}},
\label{eq:CRB}
\end{equation}
and, converting to field in unit bandwidth ($T'=1$~s),
\begin{equation}
\NEF_{\mathrm{SQL}} = \frac{\hbar}{\dip_{\mathrm{RF}}\sqrt{N\,T_2}}.
\label{eq:SQL}
\end{equation}
Note that $N$ counts every atom in the probed volume. Some authors count only
the atoms that participate in the measurement or write the limit with $h$
in place of $\hbar$, and Sec.~\ref{sec:budget} compares the conventions.

The ratio of the operating floor to this limit shows where the sensitivity
is lost. The
ratio of the shot-noise floor \eqref{eq:NEFpsn}, at the operating slope
$\mathcal{S}\kappa_T$ of \eqref{eq:kT}, to \eqref{eq:SQL} does not contain
the dipole,
\begin{equation}
\mathcal{R} \equiv \frac{\NEF_{\mathrm{PSN}}}{\NEF_{\mathrm{SQL}}}
= \frac{2\pi W_pT_2}{\mathcal{S}\,\kappa_s\,\CEIT\,\tfrac12\sin\theta_d\,(\lambda_c/\lambda_p)}
\sqrt{\frac{2N}{N_{\mathrm{ph}}}},
\label{eq:gap}
\end{equation}
with $N_{\mathrm{ph}}=\eta\Pdet T_2/h\nu_p$ the detected photons per
coherence time. Four factors make up the gap. The width of the window in
units of the coherence time contributes $2\pi W_pT_2$, the slope geometry
$[\kappa_s\CEIT\tfrac12\sin\theta_d(\lambda_c/\lambda_p)]^{-1}$, probe
saturation $1/\mathcal{S}$, and the atom--photon ratio
$\sqrt{2N/N_{\mathrm{ph}}}$. The gap closes only with a narrower window,
near-unit contrast, and a projection readout
\cite{meyer_assessment_2020,santamaria-botello_comparison_2022,kumar_atom-based_2017,ding_enhanced_2022}.
The absorptive readout of ladder EIT cannot reach the limit, because of its
absorption loss \cite{meyer_optimal_2021}. A heterodyne electrometer with laser-cooled atoms
has reached a factor of 2.6 above its projection limit
\cite{tu_approaching_2024}.

The blackbody field of the environment sets a floor that is independent of
the sensor. From the Planck energy density
$w_\nu=(8\pi h\nu^3/c^3)/(e^{h\nu/k_BT_{\mathrm{amb}}}-1)$ at the ambient
temperature $T_{\mathrm{amb}}$ and
$w=\varepsilon_0\langle E^2\rangle$, the one-sided density of the squared
field is
\begin{equation}
S_E(\nu) = \frac{w_\nu}{\varepsilon_0}
\;\xrightarrow[h\nu\ll k_BT_{\mathrm{amb}}]{}\;
\frac{8\pi\nu^2 k_BT_{\mathrm{amb}}}{\varepsilon_0 c^3},
\label{eq:planck}
\end{equation}
so a sensor reading one Cartesian component sees
\begin{equation}
\NEF_{\mathrm{th}} = \sqrt{\frac{S_E}{3}}
= \sqrt{\frac{8\pi k_B}{3\varepsilon_0 c^3}}\;\nu\sqrt{T_{\mathrm{amb}}}.
\label{eq:thermal}
\end{equation}
This is the field-domain Nyquist theorem \cite{nyquist_thermal_1928} and the
thermal limit of the Callen--Welton expression of
Ref.~\cite{santamaria-botello_comparison_2022}, whose microwave-vacuum term
overtakes it only above $\approx$4~THz at room temperature. In
superheterodyne operation the thermal field of both sidebands folds onto
the IF, which raises this floor by $\sqrt2$ for a single-sideband signal.
The thermal floor is distinct
from the atomic projection limit
\cite{itano_quantum_1993,wineland_spin_1992}. For any floor, the minimum
detectable signal in a bandwidth $B$ is $\mathrm{MDS}=\NEF\sqrt{B}$.

\subsection{Distortion}
\label{sec:nonlin}

The useful range of the receiver is bounded from above by distortion, which
the higher field derivatives of $T$ describe. The transduction from the
fields at the cell to the transmitted power is a cascade of memoryless
stages, each with its own nonlinearity. The beat between the local
oscillator and the signal forms the amplitude seen by the atoms. That
amplitude moves the dressed feature. The absorption
lineshape converts the displacement into optical depth, and the
exponential of the Beer--Lambert law, $T=e^{-d_0\bar A}$, converts the
optical depth into transmission. The treatment is the standard perturbative
expansion of a memoryless transfer characteristic
\cite{boyd_nonlinear_2020,razavi_rf_2012}, applied stage by stage.

The first stage is the beat \eqref{eq:beat}. Expanded to second order in the
complex signal envelope $\tilde e(t)$, referred to the local oscillator, the
amplitude seen by the atoms is
\begin{equation}
\big|E_L+\tilde e(t)\big| = E_L+\mathrm{Re}\,\tilde e
+\frac{(\mathrm{Im}\,\tilde e)^2}{2E_L}
+\mathcal{O}\Big(\frac{|\tilde e|^3}{E_L^2}\Big).
\label{eq:envelope}
\end{equation}
The in-phase component passes linearly, and the quadrature component
appears at second order. The mixing with the local oscillator that
linearizes the receiver also mixes the signal with itself.

The second stage is linear. The dressed energies \eqref{eq:Epm} contain the
resonant coupling of the Rydberg pair without approximation, so the probed
feature moves by $\tfrac12\kappa_\nu$ per unit field \eqref{eq:singleslope}.
Off-resonant couplings to other Rydberg levels add a quadratic ac-Stark
shift, which for the pair of the design example is about 2~kHz at 0.3~V/m,
three orders of
magnitude below the width of the window, with a second-order intercept near
460~V/m (Appendix~\ref{app:stark}). It is neglected below.

The absorption lineshape and the exponential are best treated together.
About the bias, the logarithm of the
transmission is the absorption lineshape seen through the linear map,
$\ln[T(E_L+\delta E)/T(E_L)]=\kappa_T\,\delta E+\tfrac12g_2\,\delta E^2
+\tfrac16g_3\,\delta E^3+\cdots$, with
$g_n=-d_0\,\partial^n\bar A/\partial E^n$ and $\delta E$ the excursion of
the amplitude from $E_L$. The superheterodyne bias is the
inflection of the absorption, $g_2=0$, where $\kappa_T$ is maximal, and
exponentiating gives
\begin{equation}
\frac{T(E_L+\delta E)}{T(E_L)} = 1+\kappa_T\,\delta E
+\tfrac12\kappa_T^2\,\delta E^2
+\tfrac16\big(g_3+\kappa_T^3\big)\delta E^3+\mathcal{O}(\delta E^4).
\label{eq:Texpand}
\end{equation}
At the bias the absorption lineshape contributes only odd orders, and the
exponential adds the even-order term $\tfrac12\kappa_T^2$. Its cubic term,
$\kappa_T^3$, is three orders of magnitude smaller than $g_3$ at the design
point and is dropped.

For two equal tones $\delta E=A(\cos\omega_1t+\cos\omega_2t)$, a cubic
coefficient
$a_3$ produces the in-band product at $2\omega_1-\omega_2$ with amplitude
$\tfrac34|a_3|A^3$, against the fundamental $a_1A$ in the weakly nonlinear
regime. Their extrapolated equality defines the input intercept. With
$a_1=\kappa_T$ and $a_3=g_3/6$,
\begin{equation}
A_{\mathrm{IIP3}} = \sqrt{\frac{4a_1}{3|a_3|}}
= \sqrt{\frac{8\kappa_T}{|g_3|}}
= \frac{1}{\nu_1}\sqrt{8\,\big|f'/f'''\big|},
\label{eq:AIIP3def}
\end{equation}
where $f$ is the shape of the absorption feature and
$\nu_1=\tfrac12\kappa_\nu$ the displacement slope. The natural coordinate of
$f$ is the detuning in units of the half-width of the RF-off window,
$y=(\nu-\nu_0)/(W_p/2)$. Since
the feature moves by $\tfrac12\kappa_\nu$ per unit field, one unit of $y$
corresponds to the field
\begin{equation}
\Delta E_\Gamma = \frac{W_p}{\kappa_\nu},
\label{eq:dEGamma}
\end{equation}
which changes the splitting by one full width of the window, and
\begin{equation}
A_{\mathrm{IIP3}} = \beta\,\Delta E_\Gamma,
\qquad
\beta = \sqrt{8\,\big|f'/f'''\big|}\;\Big|_{y_b},
\label{eq:beta}
\end{equation}
with the derivatives taken with respect to $y$ at the bias $y_b$. The shape
constant $\beta$ plays for distortion the role that $\kappa_s$ plays for the
slope. For a Lorentzian feature, $f(y)=1/(1+y^2)$ is biased at its
inflection $y_b=1/\sqrt3$, where
\begin{equation}
f'(y_b) = -\frac{2y_b}{(1+y_b^2)^2} = -\frac{9}{8\sqrt{3}},
\qquad
f'''(y_b) = \frac{24y_b(1-y_b^2)}{(1+y_b^2)^4} = \frac{81}{16\sqrt{3}},
\label{eq:fderivs}
\end{equation}
so $|f'/f'''|=2/9$ and $\beta=4/3$. The same construction applied
to the dressed feature of the warm vapor \eqref{eq:dressedflat} gives
$\beta=1.37$ at the design point, and $\beta$ is unchanged for the
differential readout \eqref{eq:balanced}, in which each feature still moves
at $\tfrac12\kappa_\nu$.

Single-tone compression follows from the same cubic. The fundamental of
$a_1\,\delta E+a_3\,\delta E^3$ for $\delta E=A\cos\omega t$ is
$a_1A+\tfrac34a_3A^3$, and for the
compressive sign the 1-dB point solves
$20\log_{10}[1-\tfrac34|a_3/a_1|A_{1\mathrm{dB}}^2]=-1$, so
\begin{equation}
\frac{A_{1\mathrm{dB}}}{A_{\mathrm{IIP3}}}
= \sqrt{1-10^{-1/20}} = 0.330,
\label{eq:A1dB}
\end{equation}
the 9.6-dB back-off of any memoryless-cubic front end
\cite{razavi_rf_2012,pozar_microwave_2012}.

Even-order distortion comes from the beat and from the exponential. For two
tones of amplitude $A$, the difference-frequency term of \eqref{eq:envelope}
has amplitude $A^2/2E_L$ and passes through the slope $\kappa_T$, while the
quadratic term of \eqref{eq:Texpand} adds $\tfrac12\kappa_T^2A^2$ at the same
frequency. Equating their sum with the fundamental $\kappa_TA$ gives the
intercept of the difference-frequency product,
\begin{equation}
A_{\mathrm{IIP2}} = \frac{2E_L}{|1+\kappa_TE_L|},
\label{eq:IIP2env}
\end{equation}
with $\kappa_T$ signed. On a flank where the transmission rises with field
the two terms add, and where it falls they partly cancel. At the sum
frequency the beat term changes sign, so the intercept of that product is
$2E_L/|1-\kappa_TE_L|$, and the second harmonic of a single tone has twice
that intercept. For $|\kappa_T|E_L\ll1$ both two-tone intercepts are
$2E_L$, set by the beat, and on a rising flank the difference-frequency
intercept approaches the ceiling $2/|\kappa_T|$ set by the exponential as
the drive grows. The local oscillator
sets the even-order linearity of the receiver, as the LO drive of a mixer
does, which favors a strong local oscillator on the single-feature bias.
The second-order products exceed the
third-order ones for all tone amplitudes below
$A_{\mathrm{IIP3}}^2/A_{\mathrm{IIP2}}$. Both second-order products fall
outside an IF band narrower than an octave, in which case the lineshape
sets the SFDR, while a baseband or wider IF band is limited
by the second-order SFDR.

\subsection{Spurious-free dynamic range}
\label{sec:sfdr}

The intercept and the noise floor combine into the spur-free window of the
receiver through the standard construction on logarithmic axes
\cite{urick_fundamentals_2015,razavi_rf_2012}. With input and output powers
in decibels, the fundamental has unit slope,
$P_{\mathrm{fund}}=P_{\mathrm{in}}+G$, and the third-order product slope
three through the intercept,
$P_{\mathrm{im3}}=3(P_{\mathrm{in}}-P_{\mathrm{IIP3}})+(P_{\mathrm{IIP3}}+G)$.
The product reaches the output noise floor $P_N$ at
$P^\star_{\mathrm{in}}=P_{\mathrm{IIP3}}+\tfrac13(P_N-P_{\mathrm{IIP3}}-G)$,
where the fundamental stands
\begin{equation}
\SFDR = (P^\star_{\mathrm{in}}+G)-P_N
= \tfrac23\big(P_{\mathrm{IIP3}}-P_{\mathrm{MDS}}\big)
\label{eq:SFDRconstruction}
\end{equation}
above it, with $P_{\mathrm{MDS}}=P_N-G$ the input-referred floor. In field
units the powers are $20\log_{10}$ of field amplitudes. The intercept is
the peak amplitude of each tone, whereas the NEF is an rms density, so a
tone at the floor of a 1-Hz band has the amplitude
$\sqrt2\,\NEF_{\mathrm{tot}}$ and
\begin{equation}
\SFDR = \tfrac23\,\Big[20\log_{10}\!\Big(\frac{A_{\mathrm{IIP3}}}{\sqrt2\,\NEF_{\mathrm{tot}}}\Big)\Big]
\qquad [\mathrm{dB\cdot Hz^{2/3}}].
\label{eq:SFDRdef}
\end{equation}
When second-order products fall in band the second-order SFDR is
$\tfrac12\times20\log_{10}(A_{\mathrm{IIP2}}/\sqrt2\,\NEF_{\mathrm{tot}})$ in
dB$\cdot$Hz$^{1/2}$. The linear dynamic range in a bandwidth $B$ compares
amplitudes in the same way,
$20\log_{10}(A_{1\mathrm{dB}}/\sqrt2\,\NEF\sqrt{B})$.

\subsection{Noise figure and Chu limit}
\label{sec:nf}

Comparison with conventional front ends requires translating field-referred
sensitivity into the power-referred conventions of receiver engineering.
Noise figure is defined against the available power of a source at
$T_0=290$~K \cite{friis_noise_1944,haus_representation_1960}, whereas the
atoms respond to a field, so the translation requires a collecting area. A
plane wave of rms field $E$ over a reference aperture $A_{\mathrm{ref}}$
delivers
\begin{equation}
P = \frac{E^2A_{\mathrm{ref}}}{\eta_0},
\qquad
\eta_0 = \sqrt{\mu_0/\varepsilon_0} \approx 376.7~\Omega,
\label{eq:aperture}
\end{equation}
and referring $\NEF^2$ through the same aperture gives the equivalent noise
temperature and the noise figure
\begin{equation}
T_e = \frac{\NEF^2 A_{\mathrm{ref}}}{\eta_0 k_B},
\qquad
\mathrm{NF} = 10\log_{10}\Big(1+\frac{T_e}{T_0}\Big).
\label{eq:NF}
\end{equation}
The aperture is a bookkeeping convention. The atoms sample the local field,
so the field-domain signal-to-noise
ratio
\begin{equation}
\mathrm{SNR} = \Big(\frac{\ERF}{\NEF\sqrt{B}}\Big)^2
\label{eq:SNRfield}
\end{equation}
is independent of $A_{\mathrm{ref}}$, and only the power-referred quantities
inherit it. A reported Rydberg-receiver noise figure should state
its aperture, and \eqref{eq:SNRfield} should be reported alongside. Two
conventions are used below. One is a fixed physical area
($A_{\mathrm{ref}}=1$~cm$^2$,
comparable to a vapor-cell cross-section). The other is the effective
aperture of a short dipole, $3\lambda^2/8\pi$ (0.369~cm$^2$ at 17.04~GHz),
which is $3/2$ of the isotropic aperture
$\lambda^2/4\pi$ ($0.246$~cm$^2$ at 17.04~GHz). The dipole aperture suits a
sensor that reads one field component, because the thermal field
\eqref{eq:thermal} referred through it returns $T_e=T_{\mathrm{amb}}$. With
it, \eqref{eq:NF} is the conversion of
Ref.~\cite{santamaria-botello_comparison_2022}, where the sensitivity is a
field amplitude (16\,100~K for 1.25~$\mu$V\,m$^{-1}$\,Hz$^{-1/2}$ at 10~GHz,
against 32\,200~K for the same number read as an rms density). The
amplitude reading also reproduces to 0.1~dB the noise figures of 42.44 and
19.23~dB quoted in Ref.~\cite{sandidge_resonant_2024} for 23 and
1.58~$\mu$V\,m$^{-1}$\,Hz$^{-1/2}$ at 10.22~GHz. Under either aperture a
warm vapor
sits tens of decibels above a low-noise amplifier (Sec.~\ref{sec:worked}),
in the same band as an unamplified IMDD photonic link
\cite{urick_fundamentals_2015,cox_analog_2004}.

An electrically small
antenna of radius $a$ pays the Chu--Wheeler bound on its radiation quality
factor \cite{chu_physical_1948,wheeler_fundamental_1947,mclean_re-examination_1996},
\begin{equation}
Q \gtrsim \frac{1}{(ka)^3}+\frac{1}{ka},
\qquad
k = \frac{2\pi}{\lambda},
\label{eq:chu}
\end{equation}
so a matched small antenna must be narrowband, lossy, or both. The bound
counts the reactive energy stored in the near field of the lowest spherical
mode, and it binds any structure that must deliver power to a matched load
through its own resonance. The vapor cell falls outside its premise. The
readout draws no power from the RF field (the probe laser supplies the
detected energy), the cell stores no resonant RF energy, and no impedance
match exists to be band-limited. The atomic window \eqref{eq:f3dB} sets the
instantaneous bandwidth, independent of the electrical
size $ka\ll1$ of the cell. This property has no counterpart in a small
antenna, and it permits conformal and electrically small installations
\cite{cox_quantum-limited_2018,kitching_chip-scale_2018}. However, the cell
also has no resonant field gain, and the noise figures above are those of
the bare incident field.

\subsection{The receiver chain}
\label{sec:link}

An emitter of equivalent isotropically radiated power (EIRP) at range $d$
produces the power density $\mathrm{EIRP}/4\pi d^2$, and hence the rms field
\begin{equation}
\ERF = \sqrt{\frac{\eta_0}{4\pi}}\,\frac{\sqrt{\mathrm{EIRP}}}{d},
\qquad
\sqrt{\frac{\eta_0}{4\pi}} \approx 5.48~\Omega^{1/2},
\label{eq:range}
\end{equation}
which converts a transmitter and a range directly into the field at the
cell, the quantity to which every figure of merit above refers.

Field gain ahead of the cell is available without the usual noise penalty.
Because the readout draws no power and requires no impedance match, the
cell may be placed in the converging beam of a paraboloid of effective
aperture $A_e=G_d\lambda^2/4\pi$ and focal ratio $F=f/D$. Energy
conservation concentrates the captured power $I_{\mathrm{inc}}A_e$ into a
diffraction-limited focal spot $A_f\sim(\lambda F)^2$, so the available
intensity and field gains are
\begin{equation}
G_I^{\max} \simeq \frac{A_e}{A_f} \sim \frac{G_d}{4\pi F^2},
\qquad
g_E = \sqrt{G_I}.
\label{eq:reflectorgain}
\end{equation}
This is the field-gain remedy of the
photonic-link analogy \cite{urick_fundamentals_2015}, realized with a
reflector in place of an optical amplifier and available because
the readout is non-perturbing. The atoms cannot tell a concentrated field
from a stronger plane wave, so
every intrinsic figure of merit is unchanged. The incident-referred NEF
improves by $g_E$, the intercepts refer to the incident field through the
same factor, and the SFDR is unchanged while the usable window slides to
weaker fields. While this gain adds no noise, a strong in-band interferer
reaches compression
$20\log_{10}g_E$ decibels earlier in incident terms, the usual cost of
front-end gain. The gain has further
costs. The
diffraction-limited focal volume holds fewer atoms ($N^{-1/2}$ in
\eqref{eq:SQL}), and its depth, about $8\lambda F^2$ on either side of the
focus in the paraxial estimate, can be shorter than the cell, which then
averages less than the peak field. An off-focus, mode-matched placement
with the
beam waist near the cell aperture avoids both at reduced gain. The atomic
window still sets the instantaneous bandwidth, and the reflector may be
Ruze-limited by its surface
tolerance. The self-calibration \eqref{eq:Einv} returns the local field,
so recovering the incident field requires $g_E$ from an electromagnetic
model or an in-situ calibration.

Downstream of the cell the budget follows Friis, with all noise referred to
the field input. A stage $k$ [low-noise amplifier (LNA), mixer, IF amplifier] with input-referred
voltage noise density $e_{n,k}$ contributes $e_{n,k}/G_{V,1:k-1}$ through
the cumulative conversion gain ahead of it, and independent contributions
add in quadrature,
\begin{equation}
\NEF_{\mathrm{sys}}^2
= \NEF_{\mathrm{atom}}^2
+\sum_{k\ge2}\Big(\frac{e_{n,k}}{G_{V,1:k-1}}\Big)^2,
\label{eq:friisNEF}
\end{equation}
the field-domain form of the Friis cascade of noise temperatures,
$T_{e,\mathrm{sys}}=T_{e,\mathrm{atom}}+T_{e,\mathrm{LNA}}/G_{\mathrm{atom}}+\cdots$
\cite{friis_noise_1944}. The conversion gain of the atomic stage is small
unless $\Pdet$ is large, and until it is, amplifier and mixer noise are not
suppressed, which is the situation of an unamplified photonic link
\cite{urick_fundamentals_2015,cox_analog_2004}. Raising $\Pdet$ (up to the
RIN crossover \eqref{eq:Pstar}) or adding field gain ahead of the cell is
what makes the atomic stage set the system noise figure. The intercepts
cascade in the standard way,
\begin{equation}
\frac{1}{A_{\mathrm{IIP3,sys}}^2}
\approx \frac{1}{A_{\mathrm{IIP3,atom}}^2}
+\frac{G_{\mathrm{atom}}^2}{A_{\mathrm{IIP3,LNA}}^2}+\cdots,
\label{eq:cascadeIIP3}
\end{equation}
so a high-gain front end trades the modest intercept of the atom against
that of the amplifier, as in any RF chain \cite{razavi_rf_2012}.

\section{Design example}
\label{sec:worked}

The closed forms are evaluated here for one design, from atomic
constants to link budget. Every input is a property of the atom
(Appendix~\ref{app:atomic}) or a stated experimental choice.

\subsection{Operating point}
\label{sec:operating}

Consider the $^{87}$Rb system
$5S_{1/2}\!\to\!5P_{3/2}\!\to\!50D_{5/2}\!\to\!51P_{3/2}$, with probe
$\lambda_p=780$~nm, coupling $\lambda_c=480$~nm ($\lambda_c/\lambda_p=0.615$,
$r=0.625$), and RF carrier $f_{\mathrm{RF}}=17.04$~GHz. The ARC computation
\cite{sibalic_arc_2017} of Appendix~\ref{app:dipole} gives spherical dipole components between 643
and 2033~$ea_0$ for this pair. The design adopts the $\pi$ component,
\begin{equation}
\dip_{\mathrm{RF}} = 1575\,ea_0 = 1.34\times10^{-26}~\mathrm{C\,m},
\qquad
\frac{\dip_{\mathrm{RF}}}{h} = 20.2~\mathrm{MHz/(V/m)},
\label{eq:DRFnum}
\end{equation}
with the stretched $\sigma$ component (2033~$ea_0$) available for a further
29\% of responsivity under circular polarization. The AT responsivity
\eqref{eq:kappanu} is then
\begin{equation}
\kappa_\nu = \frac{\lambda_c}{\lambda_p}\frac{\dip_{\mathrm{RF}}}{h}
= 0.615\times20.2 = 12.4~\mathrm{MHz/(V/m)}.
\label{eq:kappanunum}
\end{equation}

The probe and coupling beams share a $1/e^2$ radius $w=1$~mm in a cell at
293~K, where the most probable speed is $u=237$~m/s. (The ambient thermal
field is evaluated at the same temperature, and the noise figure is referred
to the standard 290~K.) The beam radius enters the
design through the transit dephasing, the probe power at a given
Rabi frequency, the coupling power needed for a given $\Omega_c$, and the atom
number of \eqref{eq:SQL}, and the design uses the one value $w=1$~mm for
all four. Transit
contributes a full width $\sqrt{2\ln2}\,u/(\pi w)=0.089$~MHz, whose
half-width adds to a typical warm-cell collision and laser dephasing of
$2\pi\times0.3$~MHz, so $\gamma_{31}/2\pi=\gamma_{41}/2\pi=0.344$~MHz, with
$\gamma_{21}/2\pi=3$~MHz (Appendix~\ref{app:decay}). A coupling Rabi frequency
$\Omega_c/2\pi=3.5$~MHz over this beam requires 0.6--1.0~W at 480~nm,
depending on polarization \cite{sibalic_arc_2017}. The window
\eqref{eq:Aflat} then has the widths
\begin{equation}
W_p = 2.53~\mathrm{MHz},
\qquad
W_\xi = \frac{\lambda_p}{\lambda_c}\,W_p = 4.11~\mathrm{MHz},
\label{eq:GEITbudget}
\end{equation}
in a probe scan and in the two-photon coordinate, and the contrast
$\CEIT=0.1$ is reached at $d_0=0.28$ (Sec.~\ref{sec:contrast}). The probe
runs at the shot-noise-optimal strength $\Omega_p/2\pi\approx3$~MHz
($\approx0.9\,\Omega_c$, Appendix~\ref{app:numerics}), where
$\mathcal{S}\approx\tfrac12$. Over the 1-mm beam this is 13~$\mu$W incident
and $\Pdet\approx10~\mu$W detected, at quantum efficiency $\eta=0.5$. The
local oscillator is resonant ($\sin\theta_d=1$), at $E_L=0.8$~V/m
\eqref{eq:ELrule}, with single-feature readout.

\subsection{Link budget}
\label{sec:budget}

Two sets of values appear below. The closed forms give weak-probe values,
and the numerical model gives the values at the operating probe strength,
which Table~\ref{tab:worked} lists in brackets. The noise floors use the
operating slope $\mathcal{S}\kappa_T$, and the bandwidth and the intercepts
quoted in the text are the closed forms unless stated.

At the design point the lineshape constant \eqref{eq:kappasD} is
$\kappa_s=1.47$,
and the responsivity \eqref{eq:kT} is
\begin{equation}
\kappa_T
= \kappa_s\,\frac{\CEIT}{W_p}\,\frac{\sin\theta_d}{2}\,\kappa_\nu
= 1.47\times\frac{0.1}{2.53~\mathrm{MHz}}\times\frac12\times12.4~\mathrm{\frac{MHz}{V/m}}
= 0.36~(\mathrm{V/m})^{-1}
\label{eq:kTnum}
\end{equation}
in the weak-probe limit. At the operating probe strength the slope is
$\mathcal{S}\kappa_T=0.18$~(V/m)$^{-1}$, so a field of 1~mV/m modulates the transmitted power by
$1.8\times10^{-4}$. With $\Rpd=0.31$~A/W and $Z_T=100~\mathrm{k\Omega}$ the
conversion gain \eqref{eq:GV} is $G_V\approx57$~mV/(V/m). The large
transimpedance suits the low carrier power. The Johnson noise of the
feedback resistor, $\sqrt{4k_BT_0/Z_T}=0.4$~pA/$\sqrt{\mathrm{Hz}}$, stays below
the 1.0~pA/$\sqrt{\mathrm{Hz}}$ shot noise of the detected carrier and raises
the floor below by 8\%, a term that the quoted floors omit.

The shot-noise floor \eqref{eq:NEFpsn} is, with $h\nu_p=2.55\times10^{-19}$~J,
\begin{equation}
\NEF_{\mathrm{PSN}}
= \frac{1}{0.18}\sqrt{\frac{2\times2.55\times10^{-19}}{0.5\times10^{-5}}}
= 1.8\times10^{-6}~\mathrm{\frac{V/m}{\sqrt{Hz}}}
= 18~\mathrm{\frac{nV\,cm^{-1}}{\sqrt{Hz}}}
\label{eq:NEFpsnnum}
\end{equation}
(12.5 with the differential readout \eqref{eq:balanced}, at twice the
detected power). The shot-noise-equivalent RIN at this power is
$-130$~dB/Hz. A cavity-stabilized probe at $-150$~dB/Hz at the IF adds a
power-independent 1.8~nV\,cm$^{-1}$\,Hz$^{-1/2}$ \eqref{eq:NEFrin}, the RIN
crossover \eqref{eq:Pstar} then lies at $P^\star\approx1$~mW, two orders of
magnitude above the operating power, and the optical total
\eqref{eq:NEFtot} is 18~nV\,cm$^{-1}$\,Hz$^{-1/2}$. A free-running diode laser
at $-125$~dB/Hz would instead impose a floor of 31~nV\,cm$^{-1}$\,Hz$^{-1/2}$, so the intensity
spectrum of the probe at the IF is a first-order design quantity. A coupling
laser at $-125$~dB/Hz adds 0.65~nV\,cm$^{-1}$\,Hz$^{-1/2}$.
Local-oscillator amplitude noise of $S_{\mathrm{AM}}=-130$~dB/Hz (a
single-sideband level of $-133$~dBc/Hz) on the 0.8-V/m bias adds 2.5~nV\,cm$^{-1}$\,Hz$^{-1/2}$,
comparable
to the RIN term. A residual frequency noise $\delta\nu$ of
$1~\mathrm{Hz/\sqrt{Hz}}$ maps through $\tfrac12\kappa_\nu=6.2$~MHz/(V/m) to
1.6~nV\,cm$^{-1}$\,Hz$^{-1/2}$ \eqref{eq:NEFfm}. The thermal field
\eqref{eq:thermal} at 293~K and 17~GHz is 2.0 (one component) to 3.5 (all
modes), and 2.9 for one component with both sidebands folded onto the IF.
In quadrature the optical total, the local-oscillator term, the
frequency-noise term, and the folded thermal field sum to
18~nV\,cm$^{-1}$\,Hz$^{-1/2}$. This total
lies below the 55~nV\,cm$^{-1}$\,Hz$^{-1/2}$ measured with a warm-vapor
superheterodyne receiver \cite{jing_atomic_2020} and above the
12.5~nV\,cm$^{-1}$\,Hz$^{-1/2}$ reported for a free-space warm-vapor
receiver \cite{cai_sensitivity_2022}. The system-level evaluation below
adopts 55~nV\,cm$^{-1}$\,Hz$^{-1/2}$ as a working NEF, read as an rms
density. It is a measured input, and the model does not predict it. The
optical floor scales as $1/w$ at fixed probe strength, so a larger
beam lowers it at the cost of coupling power.

The coherence time available to any interrogation protocol is bounded by
the ground--Rydberg dephasing, $T_2\approx1/\gamma_{31}=0.46~\mu$s (transit
alone would allow $2w/u=8.4~\mu$s), and the probed volume $\pi w^2L$
($L=5$~cm) holds $N=3.0\times10^8$ atoms at the density that gives
$d_0=0.28$, $2\times10^{9}$~cm$^{-3}$, which is the $^{87}$Rb density of a
natural-abundance vapor at 293~K \cite{steck_rubidium_nodate}. The
projection limit \eqref{eq:SQL} is then
\begin{equation}
\NEF_{\mathrm{SQL}}
= \frac{1.05\times10^{-34}}{1.34\times10^{-26}\sqrt{3.0\times10^{8}\times0.46\times10^{-6}}}
= 6.7~\mathrm{\frac{pV\,cm^{-1}}{\sqrt{Hz}}},
\label{eq:SQLnum}
\end{equation}
and the gap \eqref{eq:gap} is
$\mathcal{R}\approx2.7\times10^3$. The window contributes
$2\pi W_pT_2\approx7$, the slope geometry 22 (25 for a Lorentzian feature),
probe saturation 2, and the atom--photon ratio 8, with
$N_{\mathrm{ph}}\approx9\times10^6$ detected photons per coherence time.
The limit depends on two conventions. Only the velocity classes resonant
with the probe take part in the measurement, a fraction
$\bar\alpha_0=1.7\%$ of the atoms, and counting those alone, in the manner
of Ref.~\cite{jing_atomic_2020}, raises the limit to
51~pV\,cm$^{-1}$\,Hz$^{-1/2}$ and lowers the gap to 350.
Reference~\cite{tu_approaching_2024} writes \eqref{eq:SQL} with $h$ in place
of $\hbar$, which raises the limit by a further $2\pi$, to
0.32~nV\,cm$^{-1}$\,Hz$^{-1/2}$, and lowers the gap to 56.

\begin{figure}[t]
\centering
\includegraphics[width=0.55\textwidth]{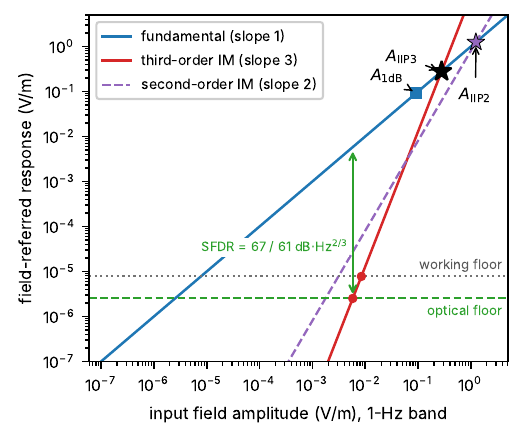}
\caption{Two-tone dynamic-range chart of the design example. The
fundamental (slope 1) and third-order
intermodulation (slope 3) responses intersect at the intercept
$A_{\mathrm{IIP3}}=0.28$~V/m, and the square marks the compression
field $A_{1\mathrm{dB}}=0.092$~V/m \eqref{eq:interceptsnum}. The second-order
product (dashed, slope 2) meets the fundamental at $A_{\mathrm{IIP2}}=1.24$~V/m
\eqref{eq:IIP2env}. The horizontal lines mark the amplitude of a tone at
unity signal-to-noise ratio in 1~Hz, $\sqrt2\,\NEF$, for the optical-noise
floor of the model (dashed) and for the working floor of
55~nV\,cm$^{-1}$\,Hz$^{-1/2}$ (dotted), which is a measured input. The
spur-free dynamic range \eqref{eq:SFDRdef}
is read where the third-order line meets each floor, 67 (optical) and 61
(working) dB$\cdot$Hz$^{2/3}$.}
\label{fig:dr}
\end{figure}

The bandwidth \eqref{eq:f3dB} and group delay \eqref{eq:taug} are
\begin{equation}
f_{3\mathrm{dB}} = \frac{W_\xi}{2} = 2.1~\mathrm{MHz},
\qquad
\tau_g(0) = \frac{1}{\pi W_\xi} = 77~\mathrm{ns},
\label{eq:f3dBnum}
\end{equation}
and the field scale of the nonlinearity is
$\Delta E_\Gamma=W_p/\kappa_\nu=0.20$~V/m \eqref{eq:dEGamma}. With
$\beta=1.37$ \eqref{eq:beta}, the intercepts and compression of
Sec.~\ref{sec:nonlin} are
\begin{equation}
A_{\mathrm{IIP3}} = \beta\,\Delta E_\Gamma = 0.28~\mathrm{V/m},
\qquad
A_{1\mathrm{dB}} = 0.092~\mathrm{V/m},
\qquad
A_{\mathrm{IIP2}} = \frac{2E_L}{1+\kappa_TE_L} = 1.24~\mathrm{V/m},
\label{eq:interceptsnum}
\end{equation}
the last from the beat and the Beer--Lambert curvature together \eqref{eq:IIP2env},
below the ceiling $2/\kappa_T=5.6$~V/m. The sum-frequency intercept is
2.2~V/m. At the third-order intercept the
second-order product lies 13~dB below the carrier, and it exceeds the
third-order product below $A_{\mathrm{IIP3}}^2/A_{\mathrm{IIP2}}\approx0.06$~V/m.
The SFDR \eqref{eq:SFDRdef} is 67~dB$\cdot$Hz$^{2/3}$ at the optical floor and
61~dB$\cdot$Hz$^{2/3}$ at the working NEF, or 57 and 52~dB$\cdot$Hz$^{1/2}$ with
in-band second-order products. The
third-order values lie well below the 110--130~dB$\cdot$Hz$^{2/3}$ reported
for microwave-photonic links \cite{marpaung_integrated_2013}, although
both devices read a small
modulation on a curved transfer characteristic. The deficit is attributed
mainly to the contrast of 0.1 and the 10~$\mu$W of detected power. The
dynamic-range ratio of Sec.~\ref{sec:designrules} grows as
$\CEIT\sqrt{\Pdet}$, so the tenfold larger modulation depth and the 1 to
10~mW of detected power of a photonic link correspond, in this scaling,
to 27 to 33~dB of SFDR. The closed-form
intercepts
are weak-probe values. At the operating probe strength the probe broadens
the window to $W_p=3.5$~MHz, which
raises them to $A_{\mathrm{IIP3}}=0.39$~V/m, $A_{1\mathrm{dB}}=0.13$~V/m, and
$A_{\mathrm{IIP2}}=1.4$~V/m in the numerical model (the last also from
\eqref{eq:IIP2env} with the operating slope). The closed form
\eqref{eq:beta} with the broadened width returns the same third-order
intercept, $\beta W_p/\kappa_\nu=0.39$~V/m. Saturation lowers the bandwidth to
1.6~MHz (Appendix~\ref{app:numerics}). The quoted SFDR is conservative by
2~dB. Figure~\ref{fig:dr} draws the two-tone construction with the
weak-probe values.

In a $B=1$~MHz channel the working MDS is
$55~\mathrm{nV\,cm^{-1}}\times\sqrt{10^6}=55~\mu$V/cm (18~$\mu$V/cm at the
optical floor), and the linear dynamic range, the amplitude at 1-dB
compression over the amplitude of a tone at the MDS, is
$20\log_{10}[0.092/(\sqrt2\times5.5\times10^{-3})]\approx21$~dB. Referred
through
\eqref{eq:NF} with $A_{\mathrm{ref}}=1$~cm$^2$, the working NEF gives a noise
figure of 33~dB (23~dB at the optical floor), and with the dipole aperture
29 and 19~dB. The thermal field of one component alone corresponds to 5.7
and 3.0~dB. The working values correspond to equivalent noise
temperatures of $2.1$--$5.8\times10^5$~K.

A prime-focus reflector of diameter $D=0.30$~m, focal ratio $F=0.4$, and
aperture efficiency $\eta_{\mathrm{ap}}=0.6$ at 17.04~GHz ($\lambda=17.6$~mm)
has
\begin{equation}
G_d = \eta_{\mathrm{ap}}\Big(\frac{\pi D}{\lambda}\Big)^2
\approx 1.7\times10^3~(32~\mathrm{dBi}),
\qquad
g_E = \sqrt{\frac{G_d}{4\pi F^2}} \approx 29,
\label{eq:dishnum}
\end{equation}
with a focal spot $A_f=\lambda^2F^2\approx0.50$~cm$^2$, larger than the
millimeter-scale beam. The focal region is $\lambda F=7.0$~mm wide and, in
the paraxial estimate, $8\lambda F^2=23$~mm deep on either side of the
focus, so a centered 5-cm cell averages 0.53 of the peak field. The values
that follow take the peak field and are upper bounds for a cell of that
length. Behind the reflector the incident-referred working NEF falls to
1.9~nV\,cm$^{-1}$\,Hz$^{-1/2}$, the MDS in 1~MHz to $\approx1.9~\mu$V/cm, and the
incident $A_{1\mathrm{dB}}$ to 3.1~mV/m. Referring signal and noise through the
same optics makes the focal spot the consistent reference aperture, and the
working noise figure referred to it is $\approx30$~dB, within $8\pi F^2/3$
(1.3~dB) of the dipole-referred value. Against the
range constant \eqref{eq:range}, a 1-W-EIRP emitter produces the working MDS
of 55~$\mu$V/cm at 1.0~km bare, or $\approx29$~km behind the reflector, and
a front-end designer continues from these numbers with
\eqref{eq:friisNEF}--\eqref{eq:cascadeIIP3}. Table~\ref{tab:worked} collects
the numbers.

\begin{table}[!tp]
\centering
\caption{Design example for $^{87}$Rb $50D_{5/2}\!\to\!51P_{3/2}$ at
17.04~GHz, with
$\CEIT=0.1$, $w=1$~mm, $\Omega_c/2\pi=3.5$~MHz, $\Omega_p/2\pi\approx3$~MHz
($\Pdet\approx10~\mu$W), $\eta=0.5$, a resonant LO at $E_L=0.8$~V/m, and
single-feature readout. Parentheses give the differential readout
\eqref{eq:balanced} at twice the detected power. Brackets give the
numerical model at
the operating probe strength (Appendix~\ref{app:numerics}) where
it differs from the closed form by more than 10\%. The working NEF is the
measured value of Ref.~\cite{jing_atomic_2020} and is an input.}
\label{tab:worked}
\small
\setlength{\tabcolsep}{3.5pt}
\begin{tabular}{llll}
\toprule
Quantity & Symbol / Eq. & Value & Units \\
\midrule
RF dipole moment ($\pi$) & $\dip_{\mathrm{RF}}$, \eqref{eq:DRFnum} & 1575 ($=1.34\times10^{-26}$~C\,m) & $ea_0$ \\
AT responsivity & $\kappa_\nu$, \eqref{eq:kappanu} & 12.4 & MHz/(V/m) \\
Window width (probe scan\,/\,two-photon) & $W_p$, $W_\xi$, \eqref{eq:GEITbudget} & 2.53\,/\,4.11 & MHz \\
Background optical depth & $d_0$, \eqref{eq:contrast} & 0.28 & --- \\
Coupling power ($\Omega_c/2\pi=3.5$~MHz) & --- & 0.6--1.0 & W \\
Detected probe power & $\Pdet$ & 10 & $\mu$W \\
Lineshape constants & $\kappa_s$, $\beta$ & 1.47, 1.37 & --- \\
Transmission responsivity, weak probe & $\kappa_T$, \eqref{eq:kTnum} & 0.36 & (V/m)$^{-1}$ \\
Transmission responsivity, operating & $\mathcal{S}\kappa_T$ & 0.18 & (V/m)$^{-1}$ \\
Conversion gain ($Z_T=100$~k$\Omega$) & $G_V$, \eqref{eq:GV} & 57 & mV/(V/m) \\
Instantaneous bandwidth & $f_{3\mathrm{dB}}$, \eqref{eq:f3dB} & 2.1\;[1.6] & MHz \\
Group delay (dc) & $\tau_g$, \eqref{eq:taug} & 77\;[168] & ns \\
Field scale of the nonlinearity & $\Delta E_\Gamma$, \eqref{eq:dEGamma} & 0.20 & V/m \\
Shot-noise NEF & \eqref{eq:NEFpsn} & 18\;(12.5) & nV\,cm$^{-1}$/$\sqrt{\mathrm{Hz}}$ \\
RIN NEF ($-150$~dB/Hz) & \eqref{eq:NEFrin} & 1.8 & nV\,cm$^{-1}$/$\sqrt{\mathrm{Hz}}$ \\
Optical-noise total & \eqref{eq:NEFtot} & 18 & nV\,cm$^{-1}$/$\sqrt{\mathrm{Hz}}$ \\
FM NEF ($1~\mathrm{Hz}/\sqrt{\mathrm{Hz}}$ residual) & \eqref{eq:NEFfm} & 1.6 & nV\,cm$^{-1}$/$\sqrt{\mathrm{Hz}}$ \\
Working NEF (measured \cite{jing_atomic_2020}) & --- & 55 & nV\,cm$^{-1}$/$\sqrt{\mathrm{Hz}}$ \\
Thermal field, 293~K (1 comp.\,/\,all) & \eqref{eq:thermal} & 2.0\,/\,3.5 & nV\,cm$^{-1}$/$\sqrt{\mathrm{Hz}}$ \\
Projection SQL ($N=3\times10^8$, $T_2=0.46~\mu$s) & \eqref{eq:SQL} & 6.7 & pV\,cm$^{-1}$/$\sqrt{\mathrm{Hz}}$ \\
Projection SQL, participating atoms & \eqref{eq:SQL} & 51 & pV\,cm$^{-1}$/$\sqrt{\mathrm{Hz}}$ \\
Third-order intercept & $A_{\mathrm{IIP3}}$, \eqref{eq:beta} & 0.28\;[0.39] & V/m \\
1-dB compression & $A_{1\mathrm{dB}}$, \eqref{eq:A1dB} & 0.092\;[0.13] & V/m \\
Second-order intercept & $A_{\mathrm{IIP2}}$, \eqref{eq:IIP2env} & 1.24\;[1.4] & V/m \\
SFDR (optical floor\,/\,working) & \eqref{eq:SFDRdef} & 67\,/\,61 & dB$\cdot$Hz$^{2/3}$ \\
MDS in $B=1$~MHz (working) & $\NEF\sqrt{B}$ & 55 & $\mu$V/cm \\
Linear DR in $B=1$~MHz (working) & $A_{1\mathrm{dB}}/\sqrt2\,\mathrm{MDS}$ & 21 & dB \\
NF, $A_{\mathrm{ref}}=1$~cm$^2$ (work./opt./thermal) & \eqref{eq:NF} & 33\,/\,23\,/\,5.7 & dB \\
NF, $A_{\mathrm{ref}}=3\lambda^2/8\pi$ & \eqref{eq:NF} & 29\,/\,19\,/\,3.0 & dB \\
Reflector field gain ($D=0.30$~m, $F=0.4$) & $g_E$, \eqref{eq:dishnum} & 29 & --- \\
Incident-referred working NEF (reflector) & --- & 1.9 & nV\,cm$^{-1}$/$\sqrt{\mathrm{Hz}}$ \\
\bottomrule
\end{tabular}
\end{table}

\subsection{Trade relations and design rules}
\label{sec:designrules}

The closed forms yield a gain--bandwidth product. The responsivity
\eqref{eq:kT}
carries $W_p$ in its denominator and the bandwidth \eqref{eq:f3dB} carries
$W_\xi=(\lambda_p/\lambda_c)W_p$ in its numerator, so their product is
independent of the width,
\begin{equation}
\kappa_T\,f_{3\mathrm{dB}}
= \frac{\kappa_s}{2}\,\CEIT\,\frac{\sin\theta_d}{2}\,\frac{\dip_{\mathrm{RF}}}{h}
= 0.74~\mathrm{MHz\,(V/m)^{-1}}
\label{eq:GBW}
\end{equation}
at the design point (0.81~MHz\,(V/m)$^{-1}$ in the numerical model, and 0.29
at the operating probe strength, where saturation lowers both factors). At
fixed contrast and lineshape, a narrower window gains responsivity only at
the expense of bandwidth. The coupling is the exception, because in the warm
vapor it also raises the contrast through $\mathcal{C}_D$ \eqref{eq:CD} and
the slope through $\Phi$ \eqref{eq:kTflat}. At fixed optical depth, raising
$\Omega_c/2\pi$ from 2 to 5~MHz raises $\kappa_Tf_{3\mathrm{dB}}$ fivefold
(Appendix~\ref{app:numerics}), so coupling power buys sensitivity and
bandwidth together, up to the power that the coupling laser can supply.
Beyond that limit, improvement requires a larger contrast (optical depth),
a larger dipole
(higher $n$, at the cost of carrier frequency, Appendix~\ref{app:atomic}),
or the differential readout.

The closed forms also yield a shot-limited dynamic-range ratio. The intercept
$A_{\mathrm{IIP3}}=\beta W_p/\kappa_\nu$ and the shot NEF, which is
$\propto W_p/\kappa_\nu$ as well, scale together, so their ratio, the
argument of the SFDR \eqref{eq:SFDRdef}, is independent of the width and of
the dipole,
\begin{equation}
\frac{A_{\mathrm{IIP3}}}{\sqrt2\,\NEF_{\mathrm{PSN}}}
= \beta\,\mathcal{S}\,\kappa_s\,\CEIT\,\frac{\sin\theta_d}{2}\,
\sqrt{\frac{\eta\Pdet}{4h\nu_p}}.
\label{eq:SFDRinv}
\end{equation}
At the shot limit the SFDR of the atomic stage is set only by the lineshape
constants, the contrast, the readout geometry, and the detected photon
rate. At fixed probe strength, $+10$~dB of SFDR costs $5.6\times$ contrast or
$32\times$ optical power (until the RIN crossover \eqref{eq:Pstar}), retuning
$n$ does not change it, and $\Omega_c$ changes it only through the contrast
and the
lineshape. The differential readout ($\sin\theta_d/2\to\sin\theta_d$) adds
$\tfrac23\times3=2$~dB of SFDR at twice the detected power, because each
feature keeps its intercept while the shot floor falls by $\sqrt2$.

These results reduce the design to a short sequence. The beam
size and the coupling power are chosen together, because the slope and the
bandwidth both rise
with $\Omega_c$ at fixed optical depth and the 480-nm
power available over the beam area sets $\Omega_c$, about 1~W for 3.5~MHz
over a 1-mm beam.
The probe runs near the shot-noise optimum
$\Omega_p\approx0.9\,\Omega_c$, with $\Pdet$ raised through the beam area
toward the
RIN crossover \eqref{eq:Pstar} and the RIN cancelled by balanced detection
against a reference tap or by the differential readout. The
local oscillator follows from the IF plan. On the single-feature bias, a local
oscillator above the plateau threshold of Sec.~\ref{sec:superhet} raises
$A_{\mathrm{IIP2}}$ \eqref{eq:IIP2env} at no cost in slope. At line center, a
local oscillator 11--14~dB weaker gives 1.5--1.6 times the slope and rejects
laser frequency noise to first order, at the cost of a second-order
intercept near $2E_L^\star$, which a sub-octave IF band tolerates. On the
single-feature bias the probe and coupling lasers are locked to a common
cavity so that the
in-band residual FM noise satisfies
$\sqrt{S_\nu}\lesssim10~\mathrm{Hz/\sqrt{Hz}}$, with the probe RIN
$\lesssim-130$~dB/Hz at the IF. These are the levels at which
\eqref{eq:NEFfm} and
\eqref{eq:NEFrin} reach the shot floor of the design example. If the
application allows a pointed aperture, reflector pre-gain
\eqref{eq:reflectorgain} lowers the incident-referred NEF, and with it the
noise temperature at a fixed reference aperture, by $20\log_{10}g_E$ with
no added noise, at the cost of
incident-referred compression
and of direct traceability.

\subsection{Validation}
\label{sec:validation}

The closed forms are tested first against the numerical model of
Appendix~\ref{app:numerics}, which averages the four-level
response over the full Maxwell--Boltzmann distribution, integrates the
equations of motion in the time domain, and replaces the weak-probe limit by
the Lindblad master equation where the probe strength matters. The closed
forms reproduce the slope to 0.05\% at equal contrast,
the third- and second-order intercepts to 0.3\%, and the bandwidth to 10\%
(Table~\ref{tab:numerics}). They fall short on the group delay, for which
the numerical model gives 116~ns, against $1/\pi W_\xi=77$~ns. The same
model locates the
line-center optimum, fixes $\mathcal{S}$, and supplies the values at the
operating probe strength.

The comparison with measurements evaluates the model at the operating
point that each experiment reports (Table~\ref{tab:compare}). The inputs
are the stated Rabi frequencies and beam sizes, atomic constants from ARC,
and the collision and laser dephasing of the design example,
$2\pi\times0.3$~MHz. Where an experiment does not
report a needed quantity, the comparison is given as a range.

\begin{table}[!tb]
\centering
\caption{The model against published measurements. In the first block the
model is evaluated with the Rabi frequencies and beam sizes that each
experiment reports, atomic constants from ARC, and the dephasing of the
design example unless stated (Appendix~\ref{app:numerics}). The second
block sets the
design example of Sec.~\ref{sec:budget} beside reported values. No
parameter is fitted. A range reflects a quantity that the experiment does
not report.}
\label{tab:compare}
\small
\setlength{\tabcolsep}{4pt}
\begin{tabular}{>{\raggedright\hangindent=1.5em\hangafter=1\arraybackslash}p{0.36\textwidth}>{\raggedright\arraybackslash}p{0.21\textwidth}>{\raggedright\arraybackslash}p{0.21\textwidth}l}
\toprule
Quantity & Model & Measured & Ref. \\
\midrule
\multicolumn{4}{l}{Model at the reported operating point}\\
\quad EIT width, Cs, $\Omega_p/\Omega_c=5.9$ (coupling scan) & 7.3~MHz & 7.9~MHz & \cite{jing_atomic_2020} \\
\quad EIT width, Cs, $\Omega_p/\Omega_c=3.4$ (probe scan) & 3.2~MHz & 3.1~MHz & \cite{artusio-glimpse_increased_2024} \\
\quad Optimum local-oscillator field & 0.32~V/m & 0.30~V/m & \cite{jing_atomic_2020} \\
\quad Half-power bandwidth, off-resonant heterodyne & 3.4--5.8~MHz ($W_\xi/2$) & 4.0~MHz & \cite{meyer_waveguide-coupled_2021} \\
\quad Double-sideband passband, $\gamma_{31}/2\pi=3$~MHz & 4.8--12~MHz ($W_\xi$) & $>10$~MHz & \cite{tang_response_2026} \\
\quad Half-amplitude offset, strong probe & 0.30~MHz (0.23--0.50) & $\approx0.4$~MHz & \cite{artusio-glimpse_increased_2024} \\
\quad Compression field, $A_{1\mathrm{dB}}/E_L$ & 0.5--1.0 & 1.0 & \cite{goncalves_nonlinearities_2026} \\
\quad Second-harmonic intercept$/E_L$ & 3.7--18 & 20 & \cite{goncalves_nonlinearities_2026} \\
\quad Two-tone third-order intercept$/E_L$ & 1.5--3.0 & 6.5 & \cite{goncalves_nonlinearities_2026} \\
\quad Shot-noise floor from the measured slope (nV\,cm$^{-1}$\,Hz$^{-1/2}$) & 17--24 & 55 (measured NEF) & \cite{jing_atomic_2020} \\
\quad Large-signal switching time & not modeled ($1/\gamma_{\mathrm{tr}}=1.8~\mu$s) & 1.0--6.1~$\mu$s & \cite{prajapati_tv_2022} \\
\midrule
\multicolumn{4}{l}{Design example beside reported values}\\
\quad Linear dynamic range in 1~Hz & 81--91~dB & 80, 90~dB & \cite{meyer_waveguide-coupled_2021,jing_atomic_2020} \\
\quad Third-order SFDR in 1~Hz & 61--67~dB & 58~dB & \cite{goncalves_nonlinearities_2026} \\
\quad Sensitivity (nV\,cm$^{-1}$\,Hz$^{-1/2}$) & 18 (optical floor) & 12.5--55 & \cite{cai_sensitivity_2022,jing_atomic_2020} \\
\midrule
\multicolumn{4}{l}{Beyond the single window (measured bandwidth, sensitivity in nV\,cm$^{-1}$\,Hz$^{-1/2}$)}\\
\quad frequency-comb probe & --- & 12~MHz, 1910 & \cite{artusio-glimpse_increased_2024} \\
\quad six-wave mixing & --- & $\pm10.2$~MHz, 62 & \cite{yang_highly_2024} \\
\quad multi-dressed states & --- & 76.8~MHz, 222.6 & \cite{yan_multi-dressed-state_2026} \\
\quad Stark-comb cell array & --- & 210~MHz, 326.6 & \cite{jiao_-demand_2026} \\
\bottomrule
\end{tabular}
\end{table}

The linewidth and the optimum local oscillator are reproduced to within
10\%. For the superheterodyne receiver of Jing \emph{et al.} (Cs
$47D_{5/2}\!\to\!48P_{3/2}$ at 6.95~GHz, $\Omega_p/2\pi=5.7$~MHz,
$\Omega_c/2\pi=0.97$~MHz), the Lindblad model gives an EIT width of 7.3~MHz
in a coupling scan, against the 7.9~MHz reported, and places the optimum
local oscillator at 0.32~V/m, where $\ORF/2\pi=0.80\,W_\xi$, against the
0.30~V/m reported \cite{jing_atomic_2020}. For the single-frequency probe of
Artusio-Glimpse \emph{et al.} (Cs, $\Omega_p/2\pi=3.7$~MHz,
$\Omega_c/2\pi\approx1.1$~MHz) it gives 3.2~MHz in a probe scan, against the
measured 3.1~MHz \cite{artusio-glimpse_increased_2024}. Both probes are
several times stronger than the coupling, so saturation sets these widths,
and the weak-probe width \eqref{eq:Aflat} underestimates them by a factor
of 2.3 to 3.3. The computed values depend on the assumed dephasing.
Between $2\pi\times0.1$ and $2\pi\times1$~MHz the two widths and the
optimum local oscillator stay within 25\% of the measurements
(Appendix~\ref{app:numerics}).

The bandwidth rule \eqref{eq:f3dB} is a weak-probe result, and the
published bandwidths probe its range of validity. Meyer \emph{et al.}
report 4.0~MHz, governed by the EIT bandwidth, for an off-resonant
heterodyne in $^{85}$Rb \cite{meyer_waveguide-coupled_2021}. The stated
coupling power and beam size correspond to a peak coupling Rabi frequency
of 9.9~MHz (ARC), for which $W_\xi/2=5.8$~MHz, and the rule gives 3.4~MHz
for 0.6 of that value. The effective Rabi frequency over the beam profile
and the magnetic sublevels is not reported, and at this coupling strength
the computed response has a shelf near the half-power level
(Appendix~\ref{app:numerics}), so the agreement is one of magnitude only.
Tang \emph{et al.} report a double-sideband passband wider than 10~MHz for
a four-level heterodyne receiver on the Cs transition of
Ref.~\cite{jing_atomic_2020}, and their density-matrix model matches the
measured response with a Rydberg dephasing of $2\pi\times3$~MHz
\cite{tang_response_2026}. With that dephasing the window \eqref{eq:Aflat}
is 4.8 to 12~MHz wide for coupling Rabi frequencies of 1 to 10~MHz, so a
double-sided passband $\approx W_\xi$ of the measured size lies within
reach of a single window. The Rabi frequencies of that experiment are not
reported, and this comparison is also one of magnitude.
In the measurement of Artusio-Glimpse \emph{et al.} the probe is 3.4 times
stronger than the coupling, and the rule fails as expected. It gives
1.15~MHz, while the Lindblad model places the offset at which the response
has fallen to half its amplitude at 0.30~MHz for the optimum local
oscillator (0.23--0.50~MHz for a local oscillator between twice and half
the optimum), against the $\approx0.4$~MHz measured
\cite{artusio-glimpse_increased_2024}.

Time-domain measurements are consistent with the single-class poles. On
the transition of the design example, Meyer \emph{et al.} found a rise
time that scales as the inverse of $\Omega_c^2/2\Gamma_e$, the coupling
term of the slow pole \eqref{eq:GammaEIT}, and a fall time bounded by the
decay of the optical coherence \cite{meyer_digital_2018}. Simons \emph{et
al.} followed the beat note of a Cs mixer to an IF of 2.5~MHz, with an
amplitude that decreases with offset \cite{simons_rydberg_2019}. However,
large-signal switching exhibits slower components that the small-signal
response does not contain. Prajapati \emph{et al.} measured rise and fall
times of 1.0 to 6.1~$\mu$s that grow with the beam width
\cite{prajapati_tv_2022}, and Bohaichuk \emph{et al.} observed a transient
shorter than 100~ns followed by an approach to steady state over about
2~$\mu$s \cite{bohaichuk_origins_2022}. These are the time scales of
population transport through the beam ($1/\gamma_{\mathrm{tr}}=1.8~\mu$s
for the 1-mm beam of the design example), and the group delay of 116 to
168~ns in the model should not be read as a switching time. The model also
omits Rydberg--Rydberg collisions and ionization, which
Ref.~\cite{bohaichuk_origins_2022} finds, with transit, to shape the
transient response.

Two measurements test the linearity. Gordon \emph{et al.} found
the mixer output linear in the signal field up to about the Autler--Townes
threshold, where their local oscillator was set \cite{gordon_weak_2019}.
Gon\c{c}alves \emph{et al.}
report, for a Cs receiver with a resonant local oscillator of 0.21~V/m, a
single-tone 1-dB compression field equal to the local-oscillator field to
within 0.1~dB, a second-order intercept of $20\,E_L$ from the IF
harmonics, and a two-tone third-order intercept of $6.5\,E_L$
\cite{goncalves_nonlinearities_2026}. On the line-center bias the model
ties every distortion scale to the local-oscillator field, in ratios that
depend on its amplitude relative to the optimum $E_L^\star$. For
$E_L=E_L^\star/2$ it gives a compression field of $1.0\,E_L$ and a
second-harmonic intercept of $18\,E_L$, close to the measured values, but a
third-order intercept of $3.0\,E_L$, a factor of two below the measurement.
For $E_L=E_L^\star$ the three values are 0.5, 3.7, and $1.5\,E_L$. The
experiment does not report the ratio of its local oscillator to the
linewidth, and its tones at 100 and 110~kHz lie above its 61-kHz IF
bandwidth, outside the quasi-static regime assumed here. The third-order
comparison remains open. The same receiver has a measured SFDR of 58~dB in
a 1-Hz band, from a two-tone intercept taken at a tone spacing of 1~MHz
\cite{goncalves_nonlinearities_2026}, against the 61 to
67~dB$\cdot$Hz$^{2/3}$ of the design example. The two receivers differ in
transition, local oscillator, and noise floor, so this agreement is
indicative.

The linear dynamic range and the sensitivity of the design example fall
within the measured ranges. The model gives 81 to 91~dB in a 1-Hz band
between the working and optical floors, against the 80 and 90~dB reported
for superheterodyne receivers in that band
\cite{meyer_waveguide-coupled_2021,jing_atomic_2020}.
The upper ends of the measured ranges are not defined at 1-dB
compression, so this agreement is also indicative. The optical floor of
18~nV\,cm$^{-1}$\,Hz$^{-1/2}$ lies between the 12.5 and
55~nV\,cm$^{-1}$\,Hz$^{-1/2}$ of warm-vapor superheterodyne receivers
\cite{cai_sensitivity_2022,jing_atomic_2020}. Because the floor
scales as $1/w$ at fixed probe strength, this comparison tests only the
bookkeeping of beam size and detected power. The slope that Jing \emph{et
al.} report gives the shot-noise floor of their receiver directly. Read as
the field per unit of optical
power at the detector, their conversion coefficient of
$7.9\times10^3$~V\,cm$^{-1}$\,W$^{-1}$ corresponds to a fractional
responsivity of 0.13~(V/m)$^{-1}$ at the roughly 10~$\mu$W on their detector,
and to a shot-noise floor of 17 to
24~nV\,cm$^{-1}$\,Hz$^{-1/2}$ for $\eta$ between 1 and 0.5. The measured
55~nV\,cm$^{-1}$\,Hz$^{-1/2}$ is a factor of 2 to 3 above that floor, which
leaves room for the technical terms of \eqref{eq:NEFtot}. The responsivity
is close to the 0.18~(V/m)$^{-1}$ of the design example. However, their
cell is optically thick (the 120-$\mu$W probe reaches the detector with
about 10~$\mu$W), so the two slopes are not directly comparable.

Receivers with larger bandwidths lie outside the single-window rule. A
frequency-comb probe reaches 12~MHz at an equivalent single-frequency
sensitivity of 1910~nV\,cm$^{-1}$\,Hz$^{-1/2}$
\cite{artusio-glimpse_increased_2024}, six-wave mixing $\pm10.2$~MHz at 62
\cite{yang_highly_2024}, multi-dressed-state engineering 76.8~MHz at 222.6
\cite{yan_multi-dressed-state_2026}, and a Stark-comb array of vapor cells
210~MHz at 326.6~nV\,cm$^{-1}$\,Hz$^{-1/2}$ \cite{jiao_-demand_2026}.

The comparisons support the cascade as a design model for the linewidth,
the optimum local oscillator, and the sensitivity. They bracket the
compression field within a factor of two, and they do not establish the
third-order intercept or the bandwidth to better than that, because the
published experiments leave the effective Rabi frequencies or the
local-oscillator level undetermined.

\section{Conclusion}
\label{sec:conclusion}

The Rydberg-atom receiver is an IMDD analog optical link whose modulator is
a vapor cell, and its link budget follows from the transmission of the
vapor. For a warm vapor that function is available in closed form
\eqref{eq:Aflat}. Its first field derivative gives the conversion gain
\eqref{eq:kT}, its width in the two-photon coordinate gives the weak-probe
bandwidth \eqref{eq:f3dB}, and its curvature, with the local-oscillator
beat, gives the intercepts \eqref{eq:beta} and \eqref{eq:IIP2env}. The noise
of the optical carrier, referred through the same slope, gives the floor
\eqref{eq:NEFtot}. A velocity-resolved numerical model reproduces the closed
forms to 1\% for the slope and the intercepts and to 10\% for the
bandwidth. It also marks their limits. The closed forms underestimate the
group delay by a third, and at the shot-noise-optimal probe strength
saturation halves the slope and lowers the bandwidth to 0.57 of half the
window width.

The formulation separates the parameters that improve the receiver from
those that only trade one figure of merit for another. Coupling power
raises sensitivity and bandwidth together \eqref{eq:GBW}, while at the shot
limit the SFDR depends only on the lineshape constants, the contrast, the
readout geometry, and the detected photon rate \eqref{eq:SFDRinv}. The gap
between the shot-noise floor and the projection limit, between 56 and
$2.7\times10^3$ according to the convention, resides in the window width,
the slope geometry, probe saturation, and the atom--photon ratio
\eqref{eq:gap}. Evaluated at published operating points, the model
reproduces the measured linewidths and the optimum
local oscillator to within 10\%, and it brackets the compression field of
a receiver whose intercepts have been measured. The measured
third-order intercept of that receiver is twice the modeled value, and this
discrepancy is unresolved.

The limiting factor in practice is the local oscillator. The
superheterodyne bias must reach the atoms as an RF field, and in laboratory
practice a horn radiates it across the cell. A receiver that broadcasts a
carrier at its own tuned frequency is not passive, and if a horn of useful
aperture must be present, terminating it in a low-noise amplifier is
generally the better receiver for passive reception. Several remedies
exist. The line-center bias of Sec.~\ref{sec:superhet} needs a local
oscillator 11--14~dB weaker. A guided structure that does not radiate into
the far field, such as a coplanar waveguide beneath the cell
\cite{meyer_waveguide-coupled_2021,wen_coplanar_1969}, or a tuned resonator
\cite{anderson_vapor-cell_2018} delivers the field locally, at the cost of
the receive geometry or of the tunability. Closed-loop and all-optical
schemes remove the RF local oscillator, at the present cost of a more
complex excitation and lower sensitivity
\cite{berweger_closed-loop_2023,schmidt_all-optical_2025}. The guided
options leave the figures of merit unchanged, because these are properties
of the local field at the atoms. The all-optical schemes replace the
EIT--AT transducer and lie outside the scope of $T$.

In its present warm-vapor form the Rydberg receiver is not a general
replacement for the antenna and low-noise amplifier. Its noise figure, its
single-window instantaneous bandwidth, and the geometry of its local
oscillator confine it to applications that depend on SI-traceable field
measurement without calibration, on non-perturbing sub-wavelength sensing,
or on calibrated spectrum monitoring from 1~GHz to 500~GHz, one
instantaneous window at a time. Closing the gap to the projection limit
requires a narrower window, higher contrast, and a projection readout, and
the local-oscillator constraints call for the guided or all-optical schemes
above. Whether these developments can be combined in one instrument remains
an outstanding problem.

\appendix

\section{Atomic physics of the transducer}
\label{app:atomic}\label{app:dipole}\label{app:stark}\label{app:decay}

A small set of atomic inputs parameterizes the transmission function. This
appendix gives the scaling of the RF carrier and the dipole moment with
principal quantum number, the angular
decomposition that yields the dipole \eqref{eq:DRFnum}, the dynamic
polarizability behind the ac-Stark shift of Sec.~\ref{sec:nonlin}, and the
physical content of the coherence-decay rates
$\gamma_{21}$, $\gamma_{31}$, and $\gamma_{41}$.

The coarse layout of the receiver follows from two scaling laws. In an
alkali atom the single valence electron in a high-$n$ state sees a
hydrogenic core, with quantum defects $\delta_{\ell j}$ correcting the
low-$\ell$ states. Energies follow the Rydberg formula with effective quantum
number $n^*=n-\delta_{\ell j}$,
$\En_n=-hcR_M/n^{*2}$ \cite{gallagher_rydberg_1994}. The carrier addressed
by the
$|3\rangle$--$|4\rangle$ transition between neighboring Rydberg states is
the local level spacing,
\begin{equation}
f_{\mathrm{RF}}
\simeq \frac{\mathrm{d}}{\mathrm{d}n}\Big(\frac{cR_M}{n^{*2}}\Big)
= \frac{2cR_M}{n^{*3}},
\label{eq:fscaling}
\end{equation}
which falls as $n^{*-3}$. For the $nD_{5/2}$--$(n+1)P_{3/2}$ transitions of
Rb, ARC gives 309~GHz at $n=20$, 17.04~GHz at $n=50$, and 2.0~GHz at
$n=100$, so the choice of $n$ is the coarse tuning of the receiver. The
orbital radius grows as
\begin{equation}
\langle r\rangle = \tfrac12 a_0\big[3n^{*2}-\ell(\ell+1)\big]
\;\propto\; n^{*2},
\label{eq:rscaling}
\end{equation}
and with it the transition dipole between neighboring Rydberg states,
$\dip\sim e\langle r\rangle\propto n^{*2}$ (reaching $\sim10^3\,ea_0$ at
$n\sim50$, three orders of magnitude beyond optical transition dipoles).
The choice of $n$ ties the dipole to the carrier frequency,
$\dip\propto f_{\mathrm{RF}}^{-2/3}$.
The design example sits at $n=50$, where the
computation below gives $f_{\mathrm{RF}}=17.04$~GHz and
$\dip_{\mathrm{RF}}=1575\,ea_0$ for the $\pi$ component, consistent with the
scaling laws.

The dipole operator is a rank-1 spherical tensor, and the Wigner--Eckart
theorem separates its geometry from its radial physics
\cite{edmonds_angular_1957},
\begin{equation}
\begin{split}
\langle n'\ell'j'm'|\,d_q\,|n\ell jm\rangle
&= (-1)^{j'-m'}
\begin{pmatrix} j' & 1 & j \\ -m' & q & m \end{pmatrix}\\
&\quad\times
\langle n'\ell'j'\Vert\,d\,\Vert n\ell j\rangle,
\end{split}
\label{eq:WE}
\end{equation}
with $q=0$ the $\pi$ ($\Delta m=0$) and $q=\pm1$ the $\sigma^\pm$ components
selected by the RF polarization relative to the optical quantization axis.
The reduced matrix element factors once more through a 6-$j$ symbol onto the
orbital element,
\begin{equation}
\begin{split}
\langle n'\ell'j'\Vert d\Vert n\ell j\rangle
&= (-1)^{j+\ell'+3/2}\sqrt{(2j+1)(2j'+1)}\\
&\quad\times
\begin{Bmatrix} \ell' & j' & 1/2 \\ j & \ell & 1 \end{Bmatrix}
\langle n'\ell'\Vert d\Vert n\ell\rangle,
\end{split}
\label{eq:sixj}
\end{equation}
and the orbital element reduces to the radial integral
$\int R_{n'\ell'}(r)\,e r\,R_{n\ell}(r)\,r^2\mathrm{d}r$, evaluated by Numerov
integration of the model-potential wavefunctions
\cite{marinescu_dispersion_1994,zimmerman_stark_1979}. The open-source ARC library
\cite{sibalic_arc_2017,robertson_arc_2021} performs the chain
\eqref{eq:WE}--\eqref{eq:sixj} numerically. For
$^{87}$Rb $50D_{5/2}\!\to\!51P_{3/2}$ it returns spherical components from
643~$ea_0$ to the stretched value 2033~$ea_0$, with the
$|m_j|=1/2$ $\pi$ component $1575\,ea_0=1.34\times10^{-26}$~C\,m adopted in
the main text \eqref{eq:DRFnum}, and it places $51P_{3/2}$ below
$50D_{5/2}$. The spread matters at the 3-dB level for the
link budget. Driving the stretched $\sigma$ component with circular
polarization and optical pumping gains 29\% of responsivity ($2033/1575$),
and an unpolarized average loses a comparable amount.

The ac-Stark shift of the pair follows from the standard second-order shift.
A field of amplitude $E$ at frequency $\omega$ shifts level $|i\rangle$ in
second order (the lowest non-vanishing Floquet order for a state with no
permanent dipole),
\begin{equation}
\Delta\En_i = -\tfrac14\,\alpha_i(\omega)\,E^2,
\qquad
\alpha_i(\omega)
= \sum_k \frac{2\,\omega_{ki}\,|\dip_{ik}|^2}{\hbar\,(\omega_{ki}^2-\omega^2)},
\label{eq:alphaAC}
\end{equation}
the dynamic (ac) polarizability, with the sum over all dipole-coupled
neighbors $k$ and $\omega_{ki}$ their transition frequencies. Rydberg states
are surrounded by a dense ladder of strongly coupled neighbors
($\dip\propto n^{*2}$, spacings $\propto n^{*-3}$), so their polarizabilities
are large, $\alpha\propto n^{*7}$ \cite{gallagher_rydberg_1994}. When a single
near-resonant neighbor dominates the sum, \eqref{eq:alphaAC} reduces to the
familiar two-level light shift
\begin{equation}
\Delta\En_{\mathrm{AC}} \approx \frac{\hbar\,\Omega_{\mathrm{RF}}^2}{4\Delta},
\label{eq:twolevelshift}
\end{equation}
with $\Delta$ the detuning from that neighbor (the same expression obtained
by expanding the dressed-state energy \eqref{eq:Epm} for
$|\dRF|\gg\ORF$). In the receiver the resonant term of the pair is
excluded from \eqref{eq:alphaAC}, because the dressing \eqref{eq:Epm}
contains it without approximation, and the remaining terms are evaluated at
the carrier
frequency. For the pair of the design example, ARC gives
$\alpha/h=+2.7\times10^3$~MHz/(V/cm)$^2$ for $50D_{5/2}$, dominated by
$49F_{7/2}$, 1.6~GHz above the carrier, and $-0.9\times10^3$~MHz/(V/cm)$^2$
for $51P_{3/2}$, dominated by $50D_{3/2}$, the fine-structure partner of the
target level, 90~MHz below the carrier. A dressed feature of a resonant
local oscillator follows the mean shift of the pair, $-\tfrac14\bar\alpha E^2$,
about 2~kHz at 0.3~V/m, so its second-order intercept,
$2(\dip_{\mathrm{RF}}/h)/|\bar\alpha/h|\approx460$~V/m, is negligible against
that of the beat (Sec.~\ref{sec:nonlin}). A neighbor within a few
linewidths of the carrier would be resonantly enhanced and would require
the same treatment as the pair itself.

The rates $\gamma_{ij}$ that damp the coherences in
\eqref{eq:eom21}--\eqref{eq:eom41} are half the sum of the population decay
rates of the two states plus any pure dephasing,
$\gamma_{ij}=\tfrac12(\Gamma_i+\Gamma_j)+\gamma_{ij}^{\mathrm{deph}}$.

The probe coherence decays at $\gamma_{21}$, which is dominated by the
radiative width of the
intermediate state. For Rb $5P_{3/2}$, $\Gamma_e=2\pi\times6.07$~MHz
\cite{steck_rubidium_nodate}, so $\gamma_{21}\approx\Gamma_e/2=2\pi\times3$~MHz. This is
the rate in the denominator of \eqref{eq:GammaEIT}, the fast polariton pole
of \eqref{eq:Hexact}, and, through the wavelength
mismatch, the additive width $r\gamma_{21}$ of the warm-vapor window
\eqref{eq:Aflat}.

The two-photon coherence decays at $\gamma_{31}$. The Rydberg state is
long-lived
(radiative plus blackbody-stimulated lifetimes of order 100~$\mu$s at
$n\sim50$ \cite{gallagher_rydberg_1994}), so $\gamma_{31}$ is
dephasing-limited. The contributions are the
transit of thermal atoms across the beams
($\sim u/w\approx2\pi\times40$--400~kHz for waists of
1--0.1~mm), Rydberg
collisions and dc Stark inhomogeneity, and the relative laser linewidth. The
residual two-photon Doppler shift enters separately, through the velocity
average \eqref{eq:Aflat}. The design example of Sec.~\ref{sec:worked} takes
$2\pi\times0.3$~MHz for the collision and laser terms, a typical warm-cell
value, to which transit through its 1-mm beam adds 0.044~MHz. Because
$T_2\sim1/\gamma_{31}$ also bounds the coherent interrogation time in the SQL
\eqref{eq:SQL}, the dephasing sets the projection floor as well as part of
the width of the window, and both enter the first factor of the gap
\eqref{eq:gap}. Cold or confined ensembles, which also remove
the residual Doppler width $r\gamma_{21}$, narrow the window and lengthen
$T_2$ together.

The RF coherence decays at $\gamma_{41}$, which has the same physical
content as $\gamma_{31}$
(two long-lived Rydberg states) and enters the innermost link of the
continued fraction \eqref{eq:rho21cf}. With $\gamma_{31}$ it sets the width
$(\gamma_{31}+\gamma_{41})/2$ of each dressed feature \eqref{eq:dressedflat},
and the design example takes $\gamma_{41}=\gamma_{31}$.

Across the tuning range the RF
transition frequency falls as $n^{*-3}$ while the RF dipole grows as
$n^{*2}$. The $5P_{3/2}$
lifetime that fixes $\gamma_{21}$ is independent of $n$, while the
$nD_{5/2}$ radiative-plus-blackbody lifetime grows as $\sim n^{*3}$ and
floors $\gamma_{31}$ (negligibly, at the kilohertz level, against warm-cell
dephasing). Propagated through \eqref{eq:Aflat} and \eqref{eq:f3dB} at
fixed coupling-laser intensity, $\Omega_c^2\propto\dip_c^2$ shrinks with
$n$, so high-$n$ (low-carrier) operating points have narrow, shallow
windows and are slow, while low-$n$ (high-carrier) points gain bandwidth at
the price of dipole moment. This is the trade of the gain--bandwidth relation
\eqref{eq:GBW}, driven from the atomic-structure side.

\section{Numerical model of the vapor}
\label{app:numerics}
\setcounter{figure}{0}
\renewcommand{\thefigure}{B\arabic{figure}}
\renewcommand{\theHfigure}{B\arabic{figure}}
\setcounter{table}{0}
\renewcommand{\thetable}{B\arabic{table}}
\renewcommand{\theHtable}{B\arabic{table}}

The closed forms of the main text assume a
Maxwell--Boltzmann weight that is flat across the homogeneous structure, a
field response that displaces one resolved dressed feature, and a weak
probe. This appendix describes a numerical model of the vapor that makes
none of these approximations and records what it gives at the design point
of
Sec.~\ref{sec:worked}, collected in Table~\ref{tab:numerics}. The model evaluates the four-level coherence
\eqref{eq:rho21cf} for each atomic velocity class and averages it over the
full Maxwell--Boltzmann distribution, linearizes and integrates the
equations of motion \eqref{eq:eom21}--\eqref{eq:eom41} for the dynamic and
two-tone tests, and replaces the weak-probe limit by the Lindblad master
equation of the four-level system where the probe strength matters. No
parameter is
adjusted to match a closed form.

\subsection*{Velocity average}
The coherence is evaluated with the substitutions of \eqref{eq:doppleravg}
and averaged over the distribution at 293~K ($u=237$~m/s,
$k_pu/2\pi=303$~MHz) by a trapezoidal sum over $1.4\times10^4$ velocity
classes, spaced by 0.02~m/s within the $\pm100$~m/s core that carries the
homogeneous structure and by 0.5~m/s beyond it, out to $5u$. The rates are
those of Sec.~\ref{sec:worked}, $\gamma_{21}/2\pi=3$~MHz,
$\Omega_c/2\pi=3.5$~MHz, and $\gamma_{31}/2\pi=\gamma_{41}/2\pi=0.344$~MHz,
with a
resonant RF transition of $\dip_{\mathrm{RF}}/h=20.15$~MHz/(V/m). The
absorption $\bar A$ is normalized to the line center of the bare
Doppler-broadened transition, so that $T=e^{-d_0\bar A}$, and $d_0$ is fixed by
requiring the RF-off contrast \eqref{eq:contrast} to equal $\CEIT=0.1$. The
responsivity $\kappa_T=-d_0\,\partial\bar A/\partial\ERF$ is obtained by
differentiating the continued fraction analytically inside the average.

\subsection*{Flat-Doppler limit}
\begin{figure}[t]
\centering
\includegraphics[width=\textwidth]{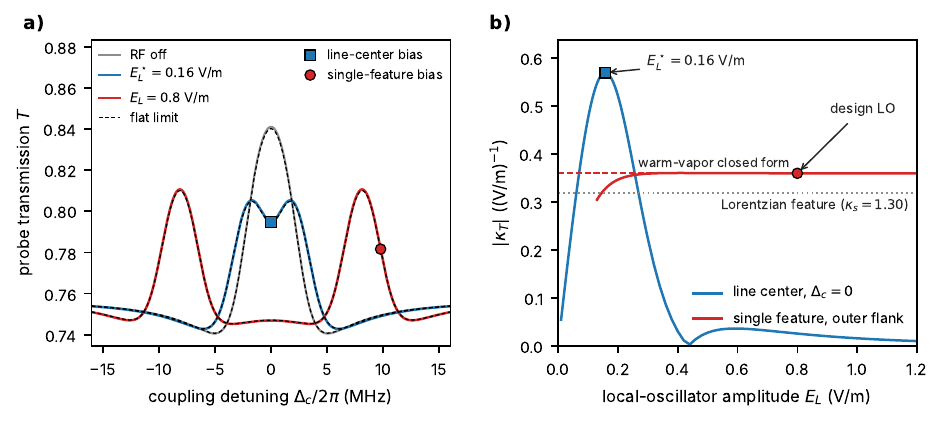}
\caption{Static transmission and the local-oscillator amplitude, from the
full Maxwell--Boltzmann average (solid) and the flat-Doppler limit
\eqref{eq:residue} (dashed). (a) Probe transmission versus coupling detuning
at $\Delta_p=0$ with the RF off (gray), at the line-center optimum
$E_L^\star=0.16$~V/m (blue), and at the design local oscillator
$E_L=0.8$~V/m (red). The markers are the two operating points. A probe scan
at $\Delta_c=0$ traces the same curves with the abscissa compressed by
$\lambda_c/\lambda_p$ \eqref{eq:xi}. (b) Responsivity versus local-oscillator
amplitude at line center (blue) and on the outer flank of the upper dressed
feature with $\Delta_c$ re-optimized at each $E_L$ (red, drawn where the
feature is defined). The dashed line is the warm-vapor closed form
\eqref{eq:kTflat}, and the dotted line is a Lorentzian feature of the same
width and contrast \eqref{eq:kappas}.}
\label{fig:numstatic}
\end{figure}

With the weight taken constant, the integral over $x=k_pv-\Delta_p$ is of a
rational function and follows by residues. With the RF off it gives
\eqref{eq:Aflat}. With the RF on, the integrand is a quadratic over a cubic in
$x$, and
\begin{equation}
\begin{split}
\bar A(\xi) &= \mathrm{Re}\Big[1+2i\sum_{\mathrm{Im}\,x_k>0}
\frac{\mathcal{Q}_2(x_k)}{\mathcal{Q}_3'(x_k)}\Big],\\
\mathcal{Q}_2(x) &= (b+irx)(c+irx)+\tfrac14\ORF^2,\qquad
\mathcal{Q}_3(x) = (\gamma_{21}-ix)\,\mathcal{Q}_2(x)+\tfrac14\Omega_c^2\,(c+irx),
\end{split}
\label{eq:residue}
\end{equation}
with $b=\gamma_{31}+i\xi$, $c=\gamma_{41}+i(\xi+\dRF)$, and $x_k$ the roots of
the cubic $\mathcal{Q}_3$. For a resolved doublet this reduces to the
dressed features
\eqref{eq:dressedflat}. Over $\pm30$~MHz of detuning and for $\ERF=0$, 0.3,
and 0.8~V/m, \eqref{eq:residue} differs from the full average by at most
$4.1\times10^{-3}$ in $\bar A$ [Fig.~\ref{fig:numstatic}(a)]. With the RF off,
the full average gives a depth $1-\bar A(0)$ of 0.378 (0.374 from
\eqref{eq:CD}), $d_0=0.278$ (0.281), and the widths $W_\xi=4.110$~MHz and
$W_p=2.529$~MHz, whose ratio is $\lambda_p/\lambda_c$ to four figures and
which \eqref{eq:Aflat} reproduces to $10^{-4}$.

\subsection*{Slope and local-oscillator amplitude}
At $E_L=0.8$~V/m the slope is maximal on the outer flank of the upper
feature, at $\Delta_c/2\pi=9.77$~MHz, where $\kappa_T=0.360$~(V/m)$^{-1}$,
and the inner flank, at $\Delta_c/2\pi=6.47$~MHz, gives $-0.355$~(V/m)$^{-1}$. There
$K_d/\Gamma_d=0.882$ and $\Phi=0.282$, and \eqref{eq:kTflat} gives
0.360~(V/m)$^{-1}$, within 0.05\% of the full average at equal contrast and
1.0\% at equal $d_0$. Over $\Omega_c/2\pi=1.5$--7~MHz and
$\gamma_{31}/2\pi=0.3$--1.5~MHz it agrees with the flat-Doppler numerics to
0.05\%, while the lineshape constant \eqref{eq:kappasD} ranges from 0.8 to
3.0. Figure~\ref{fig:numstatic}(b) shows the slope as a function of the
local-oscillator amplitude. On the outer flank it saturates once the doublet
opens. It is within 1\% of its 0.8-V/m value for $E_L\gtrsim0.26$~V/m
($\ORF/2\pi\gtrsim1.25\,W_\xi$) and varies by less than 0.2\% between 0.4
and 1.2~V/m, so the single-feature closed forms hold across this plateau.
With both lasers on resonance the local oscillator instead splits the window
symmetrically, and the slope of the falling transmission peaks at
$E_L^\star=0.158$~V/m, where $\ORF/2\pi=0.77\,W_\xi$ and the probe-scan
splitting is $0.77\,W_p$, with $\kappa_T=0.570$~(V/m)$^{-1}$, 1.58 times
(4.0~dB) the single-feature value. This line-center maximum is the global
maximum over $\Delta_c$ and $E_L$, and it is the configuration of the
superheterodyne demonstration of Ref.~\cite{jing_atomic_2020}. Over the
parameter range above, the optimum sits at
$\ORF^\star/2\pi=(0.71$--$0.84)\,W_\xi$ and exceeds the single-feature slope
by 1.4--1.9, and the rule $\Delta f_{\mathrm{AT}}\approx0.8\,W_p$ recovers the
peak slope to within 5\% over the range and to 0.2\% at the design point.

\begin{table}[!tb]
\centering
\caption{Closed forms against the numerical model for the $^{87}$Rb system
of
Sec.~\ref{sec:worked}, with $\gamma_{21}/2\pi=3$~MHz,
$\gamma_{31}/2\pi=\gamma_{41}/2\pi=0.344$~MHz, $\Omega_c/2\pi=3.5$~MHz,
$\CEIT=0.1$, and the full Maxwell--Boltzmann average at 293~K. The probe is
weak unless
stated. The closed forms refer to the single-feature bias, and their slope
at $\Omega_p/2\pi=3$~MHz includes the factor $\mathcal{S}=\tfrac12$.}
\label{tab:numerics}
\small
\begin{tabular}{lccc}
\toprule
 & Closed form & \multicolumn{2}{c}{Numerical model} \\
\cmidrule(lr){3-4}
Quantity & single feature & single feature & line center \\
\midrule
Local-oscillator amplitude $E_L$ (V/m) & 0.8 & 0.8 & 0.158 \\
Bias $\Delta_c/2\pi$ (MHz) & 9.72 & 9.77 & 0 \\
Window width $W_p$\,/\,$W_\xi$ (MHz) & 2.53\,/\,4.11 & \multicolumn{2}{c}{2.53\,/\,4.11} \\
$\kappa_T$ ((V/m)$^{-1}$) & 0.360 & 0.360 & 0.570 \\
$f_{3\mathrm{dB}}$, AM (MHz) & 2.06 & 2.26 & 2.66 \\
$f_{3\mathrm{dB}}$, upper/lower sideband (MHz) & --- & 2.68/1.82 & 2.66/2.66 \\
$\tau_g(0)$ (ns) & 77 & 116 & 105 \\
$A_{\mathrm{IIP3}}$ (V/m) & 0.279 & 0.280 & 0.235 \\
$A_{1\mathrm{dB}}$ (V/m) & 0.092 & 0.097 & 0.080 \\
$A_{\mathrm{IIP2}}$ (V/m) & 1.24 & 1.24 & 0.35 \\
First-order FM conversion & $\tfrac12\kappa_\nu$ & $\tfrac12\kappa_\nu$ & 0 \\
\midrule
\multicolumn{4}{l}{At the shot-noise-optimal probe strength, $\Omega_p/2\pi=3$~MHz:} \\
$\kappa_T$ ((V/m)$^{-1}$) & 0.18 & 0.178 & 0.272 \\
$f_{3\mathrm{dB}}$, AM (MHz) & --- & 1.62 & 1.59 \\
$\tau_g(0)$ (ns) & --- & 168 & 182 \\
$A_{\mathrm{IIP3}}$\,/\,$A_{\mathrm{IIP2}}$ (V/m) & --- & 0.39\,/\,1.4 & 0.32\,/\,0.50 \\
Shot-noise NEF (nV\,cm$^{-1}$\,Hz$^{-1/2}$) & 18 & 17 & 11 \\
\bottomrule
\end{tabular}
\end{table}

\subsection*{Line-center bias}
With a resonant local oscillator the velocity average is unchanged by the
joint reversal $\Delta_p,\Delta_c\to-\Delta_p,-\Delta_c$, so at line center
the gradient of $T$ with respect to both laser frequencies vanishes, and with
it the first-order conversion of laser frequency noise. The computed
$\partial\ln T/\partial\Delta_c$ and $\partial\ln T/\partial\Delta_p$ are zero
to numerical precision there, whereas on the outer flank the conversion
reproduces the displacement slope $\tfrac12\kappa_\nu$ of
\eqref{eq:singleslope} to 0.3\%. The local oscillator is 14~dB weaker (11~dB
at the operating probe strength, where $E_L^\star$ rises to 0.23~V/m), which
lowers the amplitude-noise floor $E_L\sqrt{S_{\mathrm{AM}}}$ by the same factor,
and the response is symmetric between the sidebands.

\begin{figure}[t]
\centering
\includegraphics[width=\textwidth]{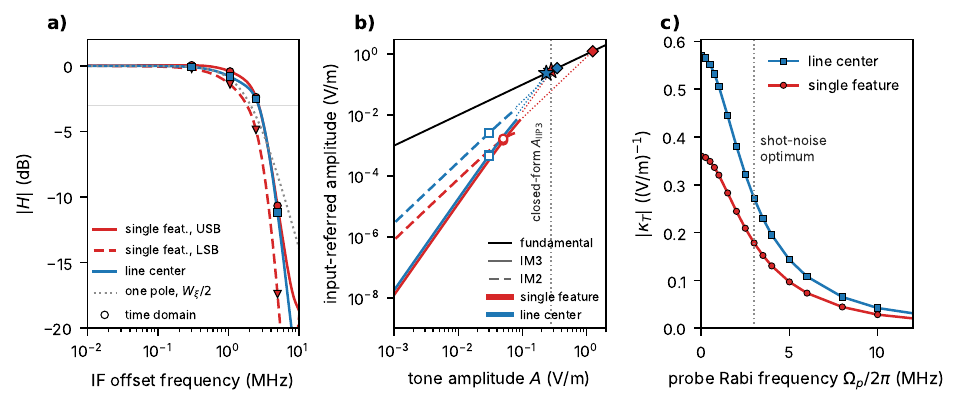}
\caption{Dynamics, distortion, and probe strength. (a) Small-signal response
at the single-feature bias for upper- (solid) and lower-sideband (dashed)
signals, and at line center (blue), with time-domain integrations (markers)
and the one-pole form \eqref{eq:Hsingle} with bandwidth $W_\xi/2$ (dotted).
(b) Two-tone test referred to the input, computed quasi-statically with the
envelope
$|E_L+\tilde e(t)|$. The curves show the third-order (solid) and
second-order (dashed) products
at the single-feature bias (red) and at line center (blue), their slope-3 and
slope-2 extrapolations (dotted), the intercepts (stars and diamonds), and
time-domain integrations (open markers). The vertical line is the
closed-form $A_{\mathrm{IIP3}}=\beta\,\Delta E_\Gamma$ \eqref{eq:beta}. (c) Responsivity versus
probe Rabi frequency from the Lindblad master equation, with the bias
re-optimized at each point. The dotted line marks the shot-noise optimum.}
\label{fig:numdyn}
\end{figure}

\subsection*{Dynamics}
The small-signal response is computed by linearizing the velocity-resolved
equations about the steady state for a signal entering as one sideband of the
local oscillator, $\ORF(t)=\Omega_L+\epsilon\,e^{\pm i\Omega t}$, and averaging
the first-order $\rho_{21}$. The double-sideband (AM) response, which
\eqref{eq:Hexact} describes, is the mean of the two single-sideband responses
[Fig.~\ref{fig:numdyn}(a)]. On the outer flank the AM half-power frequency is
2.26~MHz, within 10\% of $W_\xi/2=2.06$~MHz \eqref{eq:f3dB}, and the sidebands
differ, with 2.68~MHz for the upper and 1.82~MHz for the lower. At line
center they coincide, at 2.66~MHz. The group delay is 116~ns on the flank and
105~ns at line center, against $1/\pi W_\xi=77$~ns \eqref{eq:taug}, so the
delay--bandwidth product $\tau_g(0)\cdot2\pi f_{3\mathrm{dB}}$ is 1.65 and 1.76,
against 1 for one pole and at most $\sqrt2$ for any pair of poles. A single
velocity class at rest, with the same rates, gives 1.06~MHz and 208~ns at its
own single-feature bias, a product of 1.38, close to that of a maximally flat
pair. The three-level response \eqref{eq:Hexact}, whose poles at this
coupling are complex, $(1.67\pm1.14i)\times2\pi$~MHz, gives 1.69~MHz and
130~ns, also a product of 1.38. Both stay below $\sqrt2$, so the poles of
any one class cannot supply the excess delay of the vapor, which is
attributed to the superposition of velocity classes with different
two-photon detunings. Direct time-domain integration of the same
equations (integrating-factor fourth-order Runge--Kutta, 2-ns step) over the
classes within $\pm50$~m/s, driven by four simultaneous single-sideband tones
of 2~mV/m between 0.3 and 5~MHz, agrees with the linearized response to
$3\times10^{-4}$ in amplitude and $2\times10^{-4}$~rad in phase.

\subsection*{Distortion}
Distortion is computed quasi-statically, as $T(|E_L+\tilde e(t)|)$ with
$\tilde e(t)$ the complex two-tone envelope, which is valid for tone
frequencies well inside $f_{3\mathrm{dB}}$. A time-domain integration with
tones at 0.20 and 0.25~MHz reproduces the quasi-static fundamental,
third-order, and second-order amplitudes to within 3.3\%
[Fig.~\ref{fig:numdyn}(b)]. On the outer flank $A_{\mathrm{IIP3}}=0.280$~V/m,
against 0.279~V/m from \eqref{eq:beta}, and $A_{1\mathrm{dB}}=0.097$~V/m, a
ratio of 0.346 against the 0.330 of a pure cubic \eqref{eq:A1dB}. At line
center $A_{\mathrm{IIP3}}=0.235$~V/m and $A_{1\mathrm{dB}}=0.080$~V/m. The
second-order intercepts are 1.24~V/m on the flank and 0.35~V/m at line
center, which \eqref{eq:IIP2env} reproduces to 0.1\% at both biases, and at
the
operating probe strength it gives 1.40 and 0.50~V/m, against the computed
1.38 and 0.50~V/m. These are the intercepts of the difference-frequency
product. The sum-frequency product has intercepts of 2.24~V/m on the flank
and 0.289~V/m at line center, which $2E_L/|1-\kappa_TE_L|$ reproduces to
0.2\%, and the second harmonic of a single tone has twice the sum-frequency
intercept. At the third-order intercept the
second-order product lies 13~dB below the carrier on the flank and
3~dB at line center, so at a common detected power the line-center bias gains
1.6~dB of third-order SFDR and loses 3.6~dB of second-order SFDR. At line
center every distortion scale follows the local oscillator. For
$E_L=E_L^\star/2$ the single-tone compression field is $1.0\,E_L$, the
two-tone third-order intercept $3.0\,E_L$, and the second-harmonic
intercept $18\,E_L$, and for $E_L=E_L^\star$ the three are 0.51, 1.5, and
$3.7\,E_L$.

\subsection*{Probe strength}
The weak-probe model holds only for $\Omega_p\ll\Omega_c$, so the probe
strength is examined with the full $16\times16$ Lindblad master equation of
the four-level system,
$\dot\rho=-\tfrac{i}{\hbar}[H,\rho]+\sum_k\big(C_k\rho C_k^\dagger-\tfrac12\{C_k^\dagger C_k,\rho\}\big)$,
solved in steady state for every velocity class. The collapse operators are
radiative decay $\sqrt{\Gamma_e}\,|1\rangle\langle2|$ with $\Gamma_e/2\pi=6$~MHz,
Rydberg decay $\sqrt{\Gamma_r}\,|2\rangle\langle3|$ and
$\sqrt{\Gamma_r}\,|1\rangle\langle4|$ with $\Gamma_r/2\pi=2.3$~kHz, a common
dephasing of the Rydberg pair
$\sqrt{2\gamma_d}\,(|3\rangle\langle3|+|4\rangle\langle4|)$ with
$\gamma_d/2\pi=0.299$~MHz, and transit, represented by a reset
$\sqrt{\gamma_{\mathrm{tr}}}\,|1\rangle\langle j|$ ($j=3,4$) of the Rydberg
populations with $\gamma_{\mathrm{tr}}/2\pi=0.089$~MHz. In the weak-probe limit
these reproduce $\gamma_{21}$, $\gamma_{31}$, and $\gamma_{41}$, and at
$\Omega_p/2\pi=0.02$~MHz the master equation agrees with the weak-probe model
to $3\times10^{-5}$ in $\bar A$. As the probe strength rises, with the bias
re-optimized at each point [Fig.~\ref{fig:numdyn}(c)], the slope halves by
$\Omega_p/2\pi\approx3$~MHz and falls below 2\% of its weak-probe value by
20~MHz. At fixed beam area the detected power scales as $\Omega_p^2T$, so the
shot-noise floor \eqref{eq:NEFpsn} scales as $(\kappa_T\Omega_p\sqrt T)^{-1}$
and is minimized at $\Omega_p/2\pi\approx3$~MHz ($\approx0.9\,\Omega_c$) at
both biases. There $\kappa_T=0.178$~(V/m)$^{-1}$ on the flank and
0.272~(V/m)$^{-1}$ at line center, 0.50 and 0.48 of the weak-probe values,
which fixes $\mathcal{S}\approx\tfrac12$. The probe also broadens the RF-off
window, to $W_\xi=5.70$~MHz and $W_p=3.51$~MHz at this strength. The AM
bandwidth of 1.62~MHz on the flank is 0.57 of half the broadened width, and
the line-center optimum $E_L^\star=0.23$~V/m corresponds to a probe-scan
splitting of $0.83\,W_p$ of the broadened window. For a Gaussian beam of
$1/e^2$ radius
$w$ on the cycling transition, $P=\pi w^2I_{\mathrm{sat}}(\Omega_p/\Gamma_e)^2$
with $I_{\mathrm{sat}}=1.67$~mW/cm$^2$ \cite{steck_rubidium_nodate}, so on the
1-mm design beam the optimum corresponds to 13~$\mu$W incident and
$\Pdet=10.6~\mu$W transmitted, and the shot-noise floor with the
master-equation slopes is 17 and 11~nV\,cm$^{-1}$\,Hz$^{-1/2}$ on the flank and
at line center (12 on the flank with the differential readout
\eqref{eq:balanced}). The same model gives the first-order sensitivity to
coupling power,
$\mathrm{d}\ln T/\mathrm{d}\ln P_c=-(d_0\Omega_c/2)\,\partial\bar A/\partial\Omega_c$,
as 0.040 on the flank and 0.045 at line center for a weak probe, and as
0.021 and 0.024 at $\Omega_p/2\pi=3$~MHz. With the operating slope, a
coupling-laser RIN of $-125$~dB/Hz corresponds to
0.65~nV\,cm$^{-1}$\,Hz$^{-1/2}$ on the flank.

\subsection*{Coupling}
At fixed $d_0$, raising $\Omega_c/2\pi$ from 2 to 5~MHz raises the RF-off
contrast from 0.051 to 0.134, the single-feature slope from 0.178 to
0.486~(V/m)$^{-1}$, and the AM bandwidth from 1.60 to 3.02~MHz, so the
gain--bandwidth product \eqref{eq:GBW} grows fivefold, from 0.29 to
1.47~MHz\,(V/m)$^{-1}$. At line center it grows from 0.55 to
2.76~MHz\,(V/m)$^{-1}$. Over this range the single-feature bandwidth lies
7\% to 12\% above $W_\xi/2$, and the line-center bandwidth 22\% to 35\%
above it. At $\Omega_c/2\pi=7$~MHz the single-feature response
develops a shelf near the half-power level, which makes its $f_{3\mathrm{dB}}$
ill-conditioned, while the line-center product rises further, to
4.6~MHz\,(V/m)$^{-1}$.

\subsection*{Published operating points}
The comparison of Sec.~\ref{sec:validation} evaluates the same model with
the constants of each experiment. The wavelengths, the natural width of the
intermediate state, and the thermal speed are those of the species and the
cell temperature, the transit rate follows from the stated beam size as in
Sec.~\ref{sec:operating}, the collision and laser dephasing is kept at
$2\pi\times0.3$~MHz, and the Rydberg dipoles and Rabi frequencies come from
ARC. For Ref.~\cite{jing_atomic_2020} (Cs $47D_{5/2}\!\to\!48P_{3/2}$,
$\Omega_p/2\pi=5.7$~MHz, $\Omega_c/2\pi=0.97$~MHz, probe beam 1.70~mm in
diameter) ARC gives $\dip_{\mathrm{RF}}=1443\,ea_0$ for the $\pi$ component.
The master equation gives an RF-off window of 7.35~MHz in a coupling scan
and 4.39~MHz in a probe scan, against a weak-probe $W_\xi$ of 2.24~MHz, and
the line-center slope peaks at $E_L^\star=0.318$~V/m, where
$\ORF/2\pi=0.80\,W_\xi$. For the single-frequency probe of
Ref.~\cite{artusio-glimpse_increased_2024} (Cs, $\Omega_p/2\pi=3.7$~MHz,
$\Omega_c/2\pi=1.1$~MHz, beams of 1-mm full width at half maximum) the
window is 5.30~MHz in a coupling scan and 3.17~MHz in a probe scan, against
a weak-probe $W_\xi$ of 2.30~MHz. The line-center slope peaks at
$\ORF/2\pi=4.3$~MHz, and the AM response there falls to half power at
0.16~MHz and to half amplitude at 0.30~MHz, with a group delay of
0.92~$\mu$s. Halving or doubling the local oscillator moves the
half-amplitude point to 0.50 or 0.23~MHz. With the dephasing set to
$2\pi\times0.1$ and $2\pi\times1$~MHz, the coupling-scan width for
Ref.~\cite{jing_atomic_2020} becomes 6.95 and 8.57~MHz and its optimum
local oscillator 0.305 and 0.353~V/m, and the probe-scan width for
Ref.~\cite{artusio-glimpse_increased_2024} becomes 2.97 and 3.83~MHz. For
Ref.~\cite{meyer_waveguide-coupled_2021} ($^{85}$Rb $59D_{5/2}$, 500~mW of
coupling light in a beam of 380-$\mu$m diameter, cell at 50~$^\circ$C) ARC
gives a peak coupling Rabi frequency of 9.9~MHz for the $\pi$ transition,
and \eqref{eq:Aflat} gives $W_\xi/2=5.79$, 4.54, 3.35, and 2.30~MHz for
1.0, 0.8, 0.6, and 0.4 of that value. The weak-probe response to a
modulation of the two-photon detuning at the flank of the window reaches
half power between 2.1 and 3.3~MHz for these couplings, and for the three
larger ones it then holds a shelf between $-3$ and $-6.5$~dB out to 6 to
10~MHz. Its half-power frequency is ill-conditioned there, as it is for the
design system at $\Omega_c/2\pi=7$~MHz.
For Ref.~\cite{tang_response_2026} (the Cs transition of
Ref.~\cite{jing_atomic_2020}, probe beam 1.18~mm in diameter) the Rabi
frequencies are not reported, and the dephasing is set to the
$2\pi\times3$~MHz with which that work matches its measured response. The
window \eqref{eq:Aflat} is then 4.78, 6.11, 8.13, and 11.77~MHz wide for
$\Omega_c/2\pi=1$, 4, 6.5, and 10~MHz.

\bibliographystyle{IEEEtran}
\bibliography{rydberg}

\begin{thebibliography}{10}
\providecommand{\url}[1]{#1}
\csname url@samestyle\endcsname
\providecommand{\newblock}{\relax}
\providecommand{\bibinfo}[2]{#2}
\providecommand{\BIBentrySTDinterwordspacing}{\spaceskip=0pt\relax}
\providecommand{\BIBentryALTinterwordstretchfactor}{4}
\providecommand{\BIBentryALTinterwordspacing}{\spaceskip=\fontdimen2\font plus
\BIBentryALTinterwordstretchfactor\fontdimen3\font minus
  \fontdimen4\font\relax}
\providecommand{\BIBforeignlanguage}[2]{{%
\expandafter\ifx\csname l@#1\endcsname\relax
\typeout{** WARNING: IEEEtran.bst: No hyphenation pattern has been}%
\typeout{** loaded for the language `#1'. Using the pattern for}%
\typeout{** the default language instead.}%
\else
\language=\csname l@#1\endcsname
\fi
#2}}
\providecommand{\BIBdecl}{\relax}
\BIBdecl

\bibitem{sedlacek_microwave_2012}
J.~A. Sedlacek, A.~Schwettmann, H.~K{\"u}bler, R.~L{\"o}w, T.~Pfau, and J.~P.
  Shaffer, ``Microwave electrometry with {Rydberg} atoms in a vapour cell using
  bright atomic resonances,'' \emph{Nature Physics}, vol.~8, no.~11, pp.
  819--824, 2012.

\bibitem{fan_atom_2015}
H.~Fan, S.~Kumar, J.~Sedlacek, H.~K{\"u}bler, S.~Karimkashi, and J.~P. Shaffer,
  ``Atom based {RF} electric field sensing,'' \emph{Journal of Physics B:
  Atomic, Molecular and Optical Physics}, vol.~48, no.~20, p. 202001, 2015.

\bibitem{holloway_electric_2017}
C.~L. Holloway, M.~T. Simons, J.~A. Gordon, A.~Dienstfrey, D.~A. Anderson, and
  G.~Raithel, ``Electric field metrology for {SI} traceability: {S}ystematic
  measurement uncertainties in electromagnetically induced transparency in
  atomic vapor,'' \emph{Journal of Applied Physics}, vol. 121, no.~23, p.
  233106, 2017.

\bibitem{autler_stark_1955}
S.~H. Autler and C.~H. Townes, ``{Stark} effect in rapidly varying fields,''
  \emph{Physical Review}, vol. 100, no.~2, pp. 703--722, 1955.

\bibitem{simons_rydberg_2019}
M.~T. Simons, A.~H. Haddab, J.~A. Gordon, and C.~L. Holloway, ``A {Rydberg}
  atom-based mixer: {M}easuring the phase of a radio frequency wave,''
  \emph{Applied Physics Letters}, vol. 114, no.~11, p. 114101, 2019.

\bibitem{gordon_weak_2019}
J.~A. Gordon, M.~T. Simons, A.~H. Haddab, and C.~L. Holloway, ``Weak
  electric-field detection with sub-1 {Hz} resolution at radio frequencies
  using a {Rydberg} atom-based mixer,'' \emph{AIP Advances}, vol.~9, no.~4, p.
  045030, 2019.

\bibitem{jing_atomic_2020}
M.~Jing, Y.~Hu, J.~Ma, H.~Zhang, L.~Zhang, L.~Xiao, and S.~Jia, ``Atomic
  superheterodyne receiver based on microwave-dressed {Rydberg} spectroscopy,''
  \emph{Nature Physics}, vol.~16, no.~9, pp. 911--915, 2020.

\bibitem{holloway_broadband_2014}
C.~L. Holloway, J.~A. Gordon, S.~Jefferts, A.~Schwarzkopf, D.~A. Anderson,
  S.~A. Miller, N.~Thaicharoen, and G.~Raithel, ``Broadband {Rydberg}
  atom-based electric-field probe for {SI}-traceable, self-calibrated
  measurements,'' \emph{{IEEE} Transactions on Antennas and Propagation},
  vol.~62, no.~12, pp. 6169--6182, 2014.

\bibitem{gordon_millimeter_2014}
J.~A. Gordon, C.~L. Holloway, A.~Schwarzkopf, D.~A. Anderson, S.~Miller,
  N.~Thaicharoen, and G.~Raithel, ``Millimeter wave detection via
  {Autler}-{Townes} splitting in rubidium {Rydberg} atoms,'' \emph{Applied
  Physics Letters}, vol. 105, no.~2, p. 024104, 2014.

\bibitem{simons_using_2016}
M.~T. Simons, J.~A. Gordon, C.~L. Holloway, D.~A. Anderson, S.~A. Miller, and
  G.~Raithel, ``Using frequency detuning to improve the sensitivity of electric
  field measurements via electromagnetically induced transparency and
  {Autler}-{Townes} splitting in {Rydberg} atoms,'' \emph{Applied Physics
  Letters}, vol. 108, no.~17, p. 174101, 2016.

\bibitem{meyer_digital_2018}
D.~H. Meyer, K.~C. Cox, F.~K. Fatemi, and P.~D. Kunz, ``Digital communication
  with {Rydberg} atoms and amplitude-modulated microwave fields,''
  \emph{Applied Physics Letters}, vol. 112, no.~21, p. 211108, 2018.

\bibitem{anderson_atomic_2021}
D.~A. Anderson, R.~E. Sapiro, and G.~Raithel, ``An atomic receiver for {AM} and
  {FM} radio communication,'' \emph{{IEEE} Transactions on Antennas and
  Propagation}, vol.~69, no.~5, pp. 2455--2462, 2021.

\bibitem{urick_fundamentals_2015}
V.~J. Urick, Jr., J.~D. {McKinney}, and K.~J. Williams, \emph{Fundamentals of
  Microwave Photonics}.\hskip 1em plus 0.5em minus 0.4em\relax Hoboken, {NJ}:
  Wiley, 2015.

\bibitem{cox_analog_2004}
C.~H. Cox, III, \emph{Analog Optical Links: {T}heory and Practice}.\hskip 1em
  plus 0.5em minus 0.4em\relax Cambridge: Cambridge University Press, 2004.

\bibitem{capmany_microwave_2007}
J.~Capmany and D.~Novak, ``Microwave photonics combines two worlds,''
  \emph{Nature Photonics}, vol.~1, no.~6, pp. 319--330, 2007.

\bibitem{marpaung_integrated_2019}
D.~Marpaung, J.~Yao, and J.~Capmany, ``Integrated microwave photonics,''
  \emph{Nature Photonics}, vol.~13, no.~2, pp. 80--90, 2019.

\bibitem{chu_physical_1948}
L.~J. Chu, ``Physical limitations of omni-directional antennas,'' \emph{Journal
  of Applied Physics}, vol.~19, no.~12, pp. 1163--1175, 1948.

\bibitem{wheeler_fundamental_1947}
H.~A. Wheeler, ``Fundamental limitations of small antennas,'' \emph{Proceedings
  of the {IRE}}, vol.~35, no.~12, pp. 1479--1484, 1947.

\bibitem{cox_quantum-limited_2018}
K.~C. Cox, D.~H. Meyer, F.~K. Fatemi, and P.~D. Kunz, ``Quantum-limited atomic
  receiver in the electrically small regime,'' \emph{Physical Review Letters},
  vol. 121, no.~11, p. 110502, 2018.

\bibitem{meyer_assessment_2020}
D.~H. Meyer, Z.~A. Castillo, K.~C. Cox, and P.~D. Kunz, ``Assessment of
  {Rydberg} atoms for wideband electric field sensing,'' \emph{Journal of
  Physics B: Atomic, Molecular and Optical Physics}, vol.~53, no.~3, p. 034001,
  2020.

\bibitem{santamaria-botello_comparison_2022}
\BIBentryALTinterwordspacing
G.~Santamaria-Botello, S.~Verploegh, E.~Bottomley, and Z.~Popovic, ``Comparison
  of noise temperature of {Rydberg}-atom and electronic microwave receivers,''
  {arXiv}:2209.00908, 2022. [Online]. Available:
  \url{https://arxiv.org/abs/2209.00908}
\BIBentrySTDinterwordspacing

\bibitem{sandidge_resonant_2024}
G.~Sandidge, G.~Santamaria-Botello, E.~Bottomley, H.~Fan, and Z.~Popovi{\'c},
  ``Resonant structures for sensitivity enhancement of {Rydberg}-atom microwave
  receivers,'' \emph{{IEEE} Transactions on Microwave Theory and Techniques},
  vol.~72, no.~4, pp. 2057--2066, 2024.

\bibitem{wu_linear_2023}
F.~Wu, Q.~An, Z.~Sun, and Y.~Fu, ``Linear dynamic range of a {Rydberg}-atom
  microwave superheterodyne receiver,'' \emph{Physical Review A}, vol. 107,
  no.~4, p. 043108, 2023.

\bibitem{goncalves_nonlinearities_2026}
L.~F. Gon{\c{c}}alves, T.~Zhang, G.~Raithel, and D.~A. Anderson,
  ``Nonlinearities in atomic radio receivers: {H}armonic and intermodulation
  distortion,'' \emph{{APS} Open Science}, vol.~1, p. 000083, 2026.

\bibitem{chen_wideband_2026}
Y.~Chen, C.~Yuen, and C.~M.~S. See, ``Wideband quantum transduction for
  {Rydberg} atomic receivers using six-wave mixing,'' \emph{{IEEE} Journal on
  Selected Areas in Communications}, vol.~44, pp. 5472--5488, 2026.

\bibitem{meyer_optimal_2021}
D.~H. Meyer, C.~O'Brien, D.~P. Fahey, K.~C. Cox, and P.~D. Kunz, ``Optimal
  atomic quantum sensing using electromagnetically-induced-transparency
  readout,'' \emph{Physical Review A}, vol. 104, no.~4, p. 043103, 2021.

\bibitem{fleischhauer_electromagnetically_2005}
M.~Fleischhauer, A.~Imamoglu, and J.~P. Marangos, ``Electromagnetically induced
  transparency: {O}ptics in coherent media,'' \emph{Reviews of Modern Physics},
  vol.~77, no.~2, pp. 633--673, 2005.

\bibitem{gea-banacloche_electromagnetically_1995}
J.~Gea-Banacloche, Y.-q. Li, S.-z. Jin, and M.~Xiao, ``Electromagnetically
  induced transparency in ladder-type inhomogeneously broadened media: {T}heory
  and experiment,'' \emph{Physical Review A}, vol.~51, no.~1, pp. 576--584,
  1995.

\bibitem{artusio-glimpse_increased_2024}
\BIBentryALTinterwordspacing
A.~B. Artusio-Glimpse, D.~A. Long, S.~M. Bresler, N.~Prajapati, D.~Shylla,
  A.~P. Rotunno, M.~T. Simons, S.~Berweger, N.~Schlossberger, T.~W. {LeBrun},
  and C.~L. Holloway, ``Increased instantaneous bandwidth of {Rydberg} atom
  electrometry with an optical frequency comb probe,'' {arXiv}:2402.17942,
  2024. [Online]. Available: \url{https://arxiv.org/abs/2402.17942}
\BIBentrySTDinterwordspacing

\bibitem{yang_highly_2024}
B.~Yang, Y.~Yan, X.~Li, L.~Xiao, X.~Li, L.~Q. Chen, J.~Deng, and H.~Cheng,
  ``Highly sensitive microwave electrometry with enhanced instantaneous
  bandwidth,'' \emph{Physical Review Applied}, vol.~21, no.~3, p. L031003,
  2024.

\bibitem{yan_multi-dressed-state_2026}
\BIBentryALTinterwordspacing
Y.~Yan, B.~Yang, X.~Li, H.~Zhao, B.~Yu, J.~Deng, L.~Q. Chen, and H.~Cheng,
  ``Multi-dressed-state engineered {Rydberg} electrometry,''
  {arXiv}:2506.10541, 2026. [Online]. Available:
  \url{https://arxiv.org/abs/2506.10541}
\BIBentrySTDinterwordspacing

\bibitem{jiao_-demand_2026}
Y.~Jiao, Y.~Yin, Y.~He, J.~Hu, C.~Lu, J.~Bai, Z.~Bai, W.~Li, S.~Jia, and
  J.~Zhao, ``On-demand instantaneous bandwidth microwave receiver via scalable
  {Rydberg} vapor cell array with {Stark} comb,'' \emph{Communications
  Physics}, 2026, doi: 10.1038/s42005-026-02941-9.

\bibitem{itano_quantum_1993}
W.~M. Itano, J.~C. Bergquist, J.~J. Bollinger, J.~M. Gilligan, D.~J. Heinzen,
  F.~L. Moore, M.~G. Raizen, and D.~J. Wineland, ``Quantum projection noise:
  {P}opulation fluctuations in two-level systems,'' \emph{Physical Review A},
  vol.~47, no.~5, pp. 3554--3570, 1993.

\bibitem{degen_quantum_2017}
C.~L. Degen, F.~Reinhard, and P.~Cappellaro, ``Quantum sensing,'' \emph{Reviews
  of Modern Physics}, vol.~89, no.~3, p. 035002, 2017.

\bibitem{kumar_atom-based_2017}
S.~Kumar, H.~Fan, H.~K{\"u}bler, J.~Sheng, and J.~P. Shaffer, ``Atom-based
  sensing of weak radio frequency electric fields using homodyne readout,''
  \emph{Scientific Reports}, vol.~7, no.~1, p. 42981, 2017.

\bibitem{ding_enhanced_2022}
D.-S. Ding, Z.-K. Liu, B.-S. Shi, G.-C. Guo, K.~M{\o}lmer, and C.~S. Adams,
  ``Enhanced metrology at the critical point of a many-body {Rydberg} atomic
  system,'' \emph{Nature Physics}, vol.~18, no.~12, pp. 1447--1452, 2022.

\bibitem{tu_approaching_2024}
H.-T. Tu, K.-Y. Liao, H.-L. Wang, Y.-F. Zhu, S.-Y. Qiu, H.~Jiang, W.~Huang,
  W.~Bian, H.~Yan, and S.-L. Zhu, ``Approaching the standard quantum limit of a
  {Rydberg}-atom microwave electrometer,'' \emph{Science Advances}, vol.~10,
  no.~51, p. eads0683, 2024.

\bibitem{nyquist_thermal_1928}
H.~Nyquist, ``Thermal agitation of electric charge in conductors,''
  \emph{Physical Review}, vol.~32, no.~1, pp. 110--113, 1928.

\bibitem{wineland_spin_1992}
D.~J. Wineland, J.~J. Bollinger, W.~M. Itano, F.~L. Moore, and D.~J. Heinzen,
  ``Spin squeezing and reduced quantum noise in spectroscopy,'' \emph{Physical
  Review A}, vol.~46, no.~11, pp. R6797--R6800, 1992.

\bibitem{boyd_nonlinear_2020}
R.~W. Boyd, \emph{Nonlinear Optics}, 4th~ed.\hskip 1em plus 0.5em minus
  0.4em\relax Academic Press, 2020.

\bibitem{razavi_rf_2012}
B.~Razavi, \emph{{RF} Microelectronics}, 2nd~ed.\hskip 1em plus 0.5em minus
  0.4em\relax Upper Saddle River, {NJ}: Pearson, 2012.

\bibitem{pozar_microwave_2012}
D.~M. Pozar, \emph{Microwave Engineering}, 4th~ed.\hskip 1em plus 0.5em minus
  0.4em\relax Hoboken, {NJ}: Wiley, 2012.

\bibitem{friis_noise_1944}
H.~T. Friis, ``Noise figures of radio receivers,'' \emph{Proceedings of the
  {IRE}}, vol.~32, no.~7, pp. 419--422, 1944.

\bibitem{haus_representation_1960}
H.~A. Haus, W.~R. Atkinson, G.~M. Branch, W.~B. Davenport, Jr., W.~H. Fonger,
  W.~A. Harris, S.~W. Harrison, W.~W. {McLeod}, E.~K. Stodola, and T.~E.
  Talpey, ``Representation of noise in linear twoports,'' \emph{Proceedings of
  the {IRE}}, vol.~48, no.~1, pp. 69--74, 1960.

\bibitem{mclean_re-examination_1996}
J.~S. {McLean}, ``A re-examination of the fundamental limits on the radiation
  {Q} of electrically small antennas,'' \emph{{IEEE} Transactions on Antennas
  and Propagation}, vol.~44, no.~5, pp. 672--676, 1996.

\bibitem{kitching_chip-scale_2018}
J.~Kitching, ``Chip-scale atomic devices,'' \emph{Applied Physics Reviews},
  vol.~5, no.~3, p. 031302, 2018.

\bibitem{sibalic_arc_2017}
N.~{\v{S}}ibali{\'c}, J.~D. Pritchard, C.~S. Adams, and K.~J. Weatherill,
  ``{ARC}: {A}n open-source library for calculating properties of alkali
  {Rydberg} atoms,'' \emph{Computer Physics Communications}, vol. 220, pp.
  319--331, 2017.

\bibitem{cai_sensitivity_2022}
M.~Cai, Z.~Xu, S.~You, and H.~Liu, ``Sensitivity improvement and determination
  of {Rydberg} atom-based microwave sensor,'' \emph{Photonics}, vol.~9, no.~4,
  p. 250, 2022.

\bibitem{steck_rubidium_nodate}
D.~A. Steck, ``Rubidium 87 {D} line data,'' available online at
  \url{http://steck.us/alkalidata} (revision 2.3.4, 8 August 2025).

\bibitem{marpaung_integrated_2013}
D.~Marpaung, C.~Roeloffzen, R.~Heideman, A.~Leinse, S.~Sales, and J.~Capmany,
  ``Integrated microwave photonics,'' \emph{Laser \& Photonics Reviews},
  vol.~7, no.~4, pp. 506--538, 2013.

\bibitem{meyer_waveguide-coupled_2021}
D.~H. Meyer, P.~D. Kunz, and K.~C. Cox, ``Waveguide-coupled {Rydberg} spectrum
  analyzer from 0 to 20 {GHz},'' \emph{Physical Review Applied}, vol.~15,
  no.~1, p. 014053, 2021.

\bibitem{tang_response_2026}
Y.~Tang, S.~Wang, S.~Ren, C.~Yang, H.~Zhou, and C.~Lu, ``Response analysis of
  four-level heterodyne {Rydberg} atom receiver,'' \emph{{IEEE} Transactions on
  Antennas and Propagation}, vol.~74, no.~4, pp. 3266--3281, 2026.

\bibitem{prajapati_tv_2022}
N.~Prajapati, A.~P. Rotunno, S.~Berweger, M.~T. Simons, A.~B. Artusio-Glimpse,
  S.~D. Voran, and C.~L. Holloway, ``{TV} and video game streaming with a
  quantum receiver: {A} study on a {Rydberg} atom-based receiver's bandwidth
  and reception clarity,'' \emph{AVS Quantum Science}, vol.~4, no.~3, p.
  035001, 2022.

\bibitem{bohaichuk_origins_2022}
S.~M. Bohaichuk, D.~Booth, K.~Nickerson, H.~Tai, and J.~P. Shaffer, ``Origins
  of {Rydberg}-atom electrometer transient response and its impact on
  radio-frequency pulse sensing,'' \emph{Physical Review Applied}, vol.~18,
  no.~3, p. 034030, 2022.

\bibitem{wen_coplanar_1969}
C.~P. Wen, ``Coplanar waveguide: {A} surface strip transmission line suitable
  for nonreciprocal gyromagnetic device applications,'' \emph{{IEEE}
  Transactions on Microwave Theory and Techniques}, vol.~17, no.~12, pp.
  1087--1090, 1969.

\bibitem{anderson_vapor-cell_2018}
D.~A. Anderson, E.~G. Paradis, and G.~Raithel, ``A vapor-cell atomic sensor for
  radio-frequency field detection using a polarization-selective field
  enhancement resonator,'' \emph{Applied Physics Letters}, vol. 113, no.~7, p.
  073501, 2018.

\bibitem{berweger_closed-loop_2023}
S.~Berweger, A.~B. Artusio-Glimpse, A.~P. Rotunno, N.~Prajapati, J.~D.
  Christesen, K.~R. Moore, M.~T. Simons, and C.~L. Holloway, ``Closed-loop
  quantum interferometry for phase-resolved {Rydberg}-atom field sensing,''
  \emph{Physical Review Applied}, vol.~20, no.~5, p. 054009, 2023.

\bibitem{schmidt_all-optical_2025}
M.~Schmidt, S.~M. Bohaichuk, V.~Venu, R.~Wang, H.~K{\"u}bler, and J.~P.
  Shaffer, ``All-optical radio-frequency phase detection for {Rydberg} atom
  sensors using oscillatory dynamics,'' \emph{Physical Review Letters}, vol.
  135, no.~9, p. 093602, 2025.

\bibitem{gallagher_rydberg_1994}
T.~F. Gallagher, \emph{{Rydberg} Atoms}.\hskip 1em plus 0.5em minus 0.4em\relax
  Cambridge: Cambridge University Press, 1994.

\bibitem{edmonds_angular_1957}
A.~R. Edmonds, \emph{Angular Momentum in Quantum Mechanics}.\hskip 1em plus
  0.5em minus 0.4em\relax Princeton, {NJ}: Princeton University Press, 1957.

\bibitem{marinescu_dispersion_1994}
M.~Marinescu, H.~R. Sadeghpour, and A.~Dalgarno, ``Dispersion coefficients for
  alkali-metal dimers,'' \emph{Physical Review A}, vol.~49, no.~2, pp.
  982--988, 1994.

\bibitem{zimmerman_stark_1979}
M.~L. Zimmerman, M.~G. Littman, M.~M. Kash, and D.~Kleppner, ``{Stark}
  structure of the {Rydberg} states of alkali-metal atoms,'' \emph{Physical
  Review A}, vol.~20, no.~6, pp. 2251--2275, 1979.

\bibitem{robertson_arc_2021}
E.~J. Robertson, N.~{\v{S}}ibali{\'c}, R.~M. Potvliege, and M.~P.~A. Jones,
  ``{ARC} 3.0: {A}n expanded {Python} toolbox for atomic physics
  calculations,'' \emph{Computer Physics Communications}, vol. 261, p. 107814,
  2021.

\end{thebibliography}

\end{document}